\documentclass[prd,showpacs,amsmath,amssymb,nofootinbib,superscriptaddress]{revtex4}
\usepackage{array,mathtools,amssymb,booktabs,multirow,diagbox,hhline}
\pdfoutput=1

\usepackage{graphicx}
\usepackage{dcolumn}
\usepackage{bm}
\usepackage{bbold}

\newcommand{\Od}{{\cal O}}
\newcommand{\intT}{\int_0^\beta d\tau \int d^3 \vec{x}}

\newcommand{\mean}[1]{\left\langle{#1}\right\rangle}

\newcommand{\conds}{\langle \bar s s \rangle}
\newcommand{\condl}{\mean{\bar q q}_l}

\newcommand{\ID}{{\mathbb{1}}}

\newcommand{\eqchiral}{\!\stackrel{O(4)}{\sim}\!}
\newcommand{\equa}{\!\stackrel{U(1)_A}{\sim}\!}

\newcommand{\eqnomixp}{\!\stackrel{\theta_P=0}{\sim}\!}
\newcommand{\neqnomixp}{\!\stackrel{\theta_P=0}{\not\sim}\!}
\newcommand{\eqnomixs}{\!\stackrel{\theta_S=0}{\sim}\!}
\newcommand{\neqnomixs}{\!\stackrel{\theta_S=0}{\not\sim}\!}

\newcommand{\eqidealp}{\!\stackrel{\theta_P=\theta_P^{id}}{\sim}\!}
\newcommand{\neqidealp}{\!\stackrel{\theta_P=\theta_P^{id}}{\not\sim}\!}
\newcommand{\eqideals}{\!\stackrel{\theta_S=\theta_S^{id}}{\sim}\!}
\newcommand{\neqideals}{\!\stackrel{\theta_S=\theta_S^{id}}{\not\sim}\!}

\begin{document}

\title{Chiral and $U(1)_A$ restoration for the scalar/pseudoscalar meson nonets}
\author{A. G\'omez Nicola}
\email{gomez@fis.ucm.es}
\affiliation{Departamento de F\'{\i}sica
Te\'orica and UPARCOS. Univ. Complutense. 28040 Madrid. Spain}
\author{J. Ruiz de Elvira}
\email{elvira@itp.unibe.ch}
\affiliation{Albert Einstein Center for Fundamental Physics, Institute for Theoretical Physics,
University of Bern, Sidlerstrasse 5, CH--3012 Bern, Switzerland}

\begin{abstract}
We analyze the restoration pattern of the members of the scalar and pseudoscalar meson nonets under chiral $O(4)$ and $U(1)_A$ symmetries.
For that purpose, we exploit QCD Ward Identities (WI), which allow one to relate susceptibilities with quark condensates, as well as susceptibility differences with meson vertices. 
In addition, we consider the low-energy realization of QCD provided by $U(3)$ Chiral Perturbation Theory (ChPT) at finite temperature to perform a full analysis of the different correlators involved. 
Our analysis suggests $U(1)_A$ partner restoration if chiral symmetry partners are also degenerated.
This is also confirmed by the ChPT analysis when the light chiral limit is reached. 
Partner degeneration for the $I=1/2$ sector, the behavior of $I=0$ mixing and the temperature scaling of meson masses predicted by WI are also studied.  
Special attention is paid to the connection of our results with recent lattice analyses.

\end{abstract}

\maketitle

\section{Introduction}

The nature of chiral symmetry restoration is an essential ingredient of  the phase diagram of QCD. Chiral restoration is  realized in lattice simulations and presumably in matter formed after a Heavy Ion Collision. 
Most of its main properties are well understood. Namely, a crossover-like transition takes place in the physical case, i.e., for massive quarks and $N_f=2+1$ flavors of masses $m_u=m_d=\hat m \ll m_s$, at a transition temperature of about $T_c\sim 155$ MeV for vanishing baryon density~\cite{Aoki:2009sc,Borsanyi:2010bp,Bazavov:2011nk,Buchoff:2013nra,Bhattacharya:2014ara}. The main transition signals are the inflection of the light quark condensate and the maximum of the scalar susceptibility. As the system approaches the light chiral limit $\hat m/m_s\rightarrow 0^+$, $T_c$ decreases, the light quark condensate reduces and the scalar susceptibility peak increases at $T_c$~\cite{Ejiri:2009ac}, hence approaching the  phase transition regime characteristic of two massless flavors~\cite{Pisarski:1983ms,Smilga:1995qf}. 

In addition, the anomalous $U(1)_A$ symmetry can be asymptotically restored, driven by the vanishing of the instanton density~\cite{Gross:1980br}. 
A crucial issue with important theoretical and phenomenological consequences but which remains to be fully understood is  whether the $U(1)_A$ symmetry can be restored close to the chiral transition. 
If the answer is affirmative, the restoration pattern would be $O(4)\times U(1)_A$ instead of $SU_L(2)\times SU_R(2)\approx O(4)$ for $N_f=2$.   
Moreover, $U(1)_A$ restoration at the chiral transition not only changes the chiral pattern universality class but it also affects the order of the transition. 
It was already pointed out in~\cite{Pisarski:1983ms} that for $N_f=2$ the chiral transition would be of first order if $U(1)_A$ is  effectively restored at $T_c$ and of second order if it is not. 
This has been also confirmed by recent effective model analysis~\cite{Eser:2015pka}. The restoration of $U(1)_A$ would also affect the transition order for $N_f=3$~\cite{Pelissetto:2013hqa} as well as the behavior near the critical end point at finite temperature and baryon chemical potential~\cite{Mitter:2013fxa}. Analyses of $U(1)_A$ restoration using effective theories for $N_f=3$ have also been carried out recently~\cite{Fejos:2015xca} reaching similar conclusions. 

The particle spectrum would also be directly affected. In particular, the physical states becoming chiral partners, i.e., those that degenerate at the transition, would be different depending on the chiral pattern.
This would also have a direct consequence is the behavior of the associated susceptibilities and screening masses. 
On the one hand, within the scalar $0^{++}$ and pseudoscalar $0^{-+}$  meson nonets, if the chiral group $SU_L(2)\times SU_R(2)$ is restored, the pion and the $\sigma/f_0(500)$ are expected to degenerate~\cite{Hatsuda:1985eb,Krippa:2000jh}. On the other hand, the restoration of the $U(1)_A$ symmetry would allow the pion to be degenerated with the $a_0(980)$, i.e., the member of the scalar nonet with the same pion quantum numbers but an opposite parity.  
In this context, it is also natural to investigate the fate of the rest of the members of the scalar and pseudoscalar nonet at chiral restoration, i.e., the $K(800)$ (or $\kappa$) versus the kaon for $I=1/2$, and the $f_0(980)-f_0(500)$ pair versus the $\eta-\eta'$ for the $I=0$ octet and singlet members.  

Regarding an effective low-energy description, if chiral and $U(1)_A$ restoration happen to be close, a proper description of this regime will require the $\eta_0$ (singlet) state to be included formally as the ninth Goldstone boson~\cite{Witten:1979vv,HerreraSiklody:1996pm,Kaiser:2000gs}, which at $T=0$ relies on the large $N_c$ limit~\cite{Witten:1979vv,HerreraSiklody:1996pm,Kaiser:2000gs,otahetal}.  
In fact, there is experimental evidence of the reduction of the $\eta'$ mass in the hot medium~\cite{Csorgo:2009pa}, which also points out to $U(1)_A$ restoration and confirms the early proposal in~\cite{Kapusta:1995ww},  
where  phenomenological effects of the $\eta'$  mass reduction on the dilepton and diphoton spectra are analyzed.

However, $U(1)_A$ restoration is meant to be reached only asymptotically. Thus, it is important to clarify that by $U(1)_A$ restoration we will mean the approximate degeneration of $U(1)_A$ partners in comparison with $O(4)$ partner degeneration. The idea that $U(1)_A$ partners can degenerate in an ideal chiral restoring scenario was first suggested in~\cite{Shuryak:1993ee} and confirmed in~\cite{Cohen:1996ng} through an analysis of spectral properties of the QCD quark propagator. Nevertheless, in the real world with massive quarks, nontrivial gauge configurations make in general a nonzero $U(1)_A$ breaking to be present~\cite{Lee:1996zy} even though $U(1)_A$ partners could still be approximately degenerate. The particle spectrum at finite temperature including the $U(1)_A$ anomaly has been studied within a linear sigma model description in~\cite{Meggiolaro:2013swa} and using renormalization-group methods in~\cite{Heller:2015box}. In addition, screening and pole masses at $U(1)_A$ restoration within the NJL model are studied in~\cite{Ishii:2015ira,Ishii:2016dln}. A recent work intimately connected with our present analysis is~\cite{Azcoiti:2016zbi}, where the $O(4)$ and $U(1)_A$ transitions are studied in terms of the topological susceptibility.

The restoration of the $U(1)_A$ symmetry also affects the temperature dependence of the $\eta-\eta'$ mixing. Since the vanishing of the anomalous contribution to the $\eta'$ mass implies ideal mixing~\cite{Guo:2012ym,Guo:2015xva,Gu:2018swy}, i.e., the $\eta$ and $\eta'$ states being of pure light and strange quark content respectively, one would naturally expect that at temperatures where $U(1)_A$ is restored, the mixing angle should reach the ideal limit. 
This is indeed a nontrivial statement since the $T=0$ physical mixing angle is far from the ideal one. The ideal limit at asymptotically high temperatures has been confirmed by recent analysis within the Linear Sigma Model~\cite{Lenaghan:2000ey} and the NJL model~\cite{Ishii:2016dln}.  
 
The above aspects regarding chiral partners and patterns are also of fundamental relevance to clarify the nature of the scalar nonet, which has been a matter of debate over the recent past~\cite{Olive:2016xmw,RuizdeElvira:2010cs,Pelaez:2015qba}. Thus, the restoration pattern could help to shed light on the nature of those states when compared to the predictions based upon their $\bar q q$ assignment. 
Note that the full restoration of the chiral $SU_L(3)\times SU_R(3)\times U(1)_A$ symmetry would imply a complete degeneration of all members of the two nonets.  
Hence, it is expected that it would take place at a much higher temperature, since it requires the vanishing of the $\conds$ condensate, which has a much softer temperature dependence~\cite{Buchoff:2013nra}. 

Many of the issues described above regarding chiral and $U(1)_A$ restoration have been recently analyzed also by lattice collaborations. Nonetheless, the nature of the chiral pattern is still subject to debate. 
On the one hand, for $N_f=2+1$ flavors and nonzero quark masses, it has been  found in~\cite{Buchoff:2013nra,Bhattacharya:2014ara} that the $U(1)_A$ symmetry in terms of $\pi-a_0$ partner degeneration is restored well above $T_c$, i.e., the chiral transition temperature where $\pi-\sigma$ states degenerate.  These results are consistent with previous analysis of screening masses by the same group~\cite{Cheng:2010fe}.
Another lattice analysis based on meson screening masses pointing towards $U(1)_A$ restoration taking place above $T_c$ is~\cite{Lee:2017uvl},  for two flavors and two colors. 
On the other hand, the lattice results in~\cite{Aoki:2012yj,Cossu:2013uua,Tomiya:2016jwr} are consistent with $U(1)_A$ restoration taking place at the chiral transition or very close above it. 
These simulations are performed in the chiral limit for two flavors. In addition, in the recent analysis~\cite{Brandt:2016daq}, results compatible with $U(1)_A$ being restored at the chiral transition are also reported for two flavors and massive quarks.   The influence of  $U(1)_A$ restoration on the phase diagram, the tricritical point and the transition order has also been investigated in the lattice in~\cite{Chandrasekharan:2007up,Chandrasekharan:2010ik}, while degeneration of parity partners for nucleons in the lattice have been analyzed recently in~\cite{Aarts:2015mma}. 

Aiming to provide as much theoretical information as possible, we will carry out here a detailed analysis of chiral and $U(1)_A$ symmetry restoration for the scalar and pseudoscalar nonets based on Ward Identities (WI) and $U(3)$ Chiral Perturbation Theory (ChPT). Our present analysis extends in a nontrivial way the $SU(2)$ study performed in~\cite{Nicola:2013vma}, where WI played a crucial role to describe the degeneration of $\sigma$-$\pi$ states, and in~\cite{Nicola:2016jlj}, where $U(3)$  WI  relating quark condensates and pseudoscalar susceptibilities were derived and checked within $U(3)$ ChPT. 
In a recent work~\cite{GomezNicola:2017bhm}, we have exploited  WI to reach useful conclusions about $U(1)_A$ and chiral restoration and derived new WI connecting two and three point functions.  
Here, we will carry out  a detailed derivation of all the relevant WI and we will discuss their main consequences for chiral and $U(1)_A$ restoration. In addition, we will show that new scalar WI provide a good description of the scaling of lattice screening masses in the scalar $I=1/2$ sector, thus extending  the  analysis in~\cite{Nicola:2013vma} and~\cite{Nicola:2016jlj}. 
We will also  perform a full analysis within the framework of $U(3)$ ChPT. 
On the one hand, ChPT is needed to provide a specific realization of WI for hadronic states, since WI are formally derived from QCD, and therefore they may be subject to renormalization ambiguities. 
On the other hand, our ChPT analysis will provide support for our results in~\cite{GomezNicola:2017bhm} regarding partner degeneration. It will also allow us to analyze carefully the behavior near the chiral limit within a model-independent approach. Such model independency  is the main advantage of the ChPT framework. 

The paper is organized as follows: in Section~\ref{sec:gen} we will provide the general derivation of the WI considered. 
The consequences of one-point WI for the different isospin sectors of the two nonets will be discussed in Section~\ref{sec:twoWI}, which includes  new analyses of isospin-breaking, the role of connected and disconnected susceptibilities and screening  masses. Section~\ref{sec:threeWI} will be devoted to the analysis of two-point WI. The effective field theory analysis based on $U(3)$ ChPT will be presented in Section~\ref{sec:chpt},
where we will provide a model-independent and renormalizable hadron realization of the WI analyzed in previous sections.  
In addition, we will also analyze the ChPT predictions for the temperature behavior of the relevant observables for chiral and $U(1)_A$ restoration. 
Actually, the explicit expressions for the scalar susceptibilities in this formalism are derived here for the first time and collected in Appendix \ref{sec:app}. 
Our ChPT results, albeit limited at temperatures well below the transition, will essentially capture and confirm the main results obtained formally from the analysis of WI. 
They will be particularly useful in the chiral limit and will provide new insights on this problem for future lattice and theoretical analysis. 
Finally, in Section~\ref{sec:conc} we will present our main conclusions.

\section{General Ward Identities}
\label{sec:gen}

In order to clarify partner degeneration in terms of different symmetry restoration patterns, we consider the pseudoscalar $P^a=i\bar \psi \gamma_5 \lambda^a\psi$ and scalar $S^a=\bar\psi \lambda^a\psi$ quark bilinears, with $\psi$ a three-flavor fermion field with components $\psi_{u,d,s}$, $\lambda^{a=1,\dots 8}$ the $SU(3)$ Gell-Mann matrices and $\lambda^0=\sqrt{2/3}\,\ID$. We follow the same notation as in~\cite{Nicola:2016jlj,GomezNicola:2017bhm}.  

The relevant transformations to study the restoration of the chiral and $U(1)_A$ symmetries are those of the parity-changing $U_A(3)$ group, i.e., the  infinitesimal transformations $\delta\psi=i\alpha_A^a\frac{ \lambda_a}{2}\gamma_5\psi$ and $\delta\bar\psi=i\alpha_A^a\bar\psi\frac{\lambda_a}{2}\gamma_5$. Note that a  $SU_V(3)$ transformation would always allow one to rotate between members of the same octet, i.e., without change of parity.  
Under such axial transformations, the expectation value of an arbitrary pseudoscalar local operator $\mathcal{O_P}(x_1,\cdots,x_n)$ in terms of the transformed fields leads to the following  generic WI~\cite{Nicola:2016jlj,GomezNicola:2017bhm}  
\begin{align}
&\left\langle\frac{\delta\mathcal{O_P}(x_1,\cdots,x_n)}{\delta \alpha_A^a(x)}\right\rangle=
-\left\langle\mathcal{O_P}(x_1,\cdots,x_n)\bar\psi(x)\left\{\frac{\lambda^a}{2},\mathcal{M}\right\}\gamma_5\psi(x)\right\rangle 
+i\frac{\delta_{a0}}{\sqrt{6}}\left\langle \mathcal{O_P}(x_1,\cdots,x_n) A(x)\right\rangle,
\label{wigen}
\end{align}
where 
\begin{equation}
  A(x)=\frac{3g^2}{16\pi^2}\mbox{Tr}_c G_{\mu\nu}\tilde G^{\mu\nu},
  \label{anomaly}
\end{equation}
is the anomalous divergence of the $U(1)_A$ current~\cite{Bardeen:1969md,Adler:1969er,Fujikawa:1980eg}, 
\begin{equation} \label{anomalyeq}
  \partial_\mu J_5^\mu=2i\bar\psi\mathcal{M}\gamma_5\psi+A(x),
\end{equation}
with $J_5^\mu=\bar\psi \gamma_5\gamma^\mu\psi$. Generally speaking, applying~\eqref{wigen} to an $n$-point operator $\mathcal{O_P}$ gives an identity relating correlators of $n$ and $n+1$ points. 

In the same way, WI obtained from a isovector transformation $\delta\psi=i\alpha_V^a\frac{ \lambda_a}{2}\psi$, $\delta\bar\psi=-i\alpha_V^a\bar\psi\frac{\lambda_a}{2}$ on a scalar operator $\mathcal{O_S}$ read:
\begin{align}
&\left\langle\frac{\delta\mathcal{O_S}(x_1,\cdots,x_n)}{\delta \alpha_V^a(x)}\right\rangle=\left\langle\mathcal{O_S}(x_1,\cdots,x_n)\bar\psi(x)\left[\frac{\lambda^a}{2},\mathcal{M}\right]\psi(x)\right\rangle.
\label{wigenvec}
\end{align}

The analysis to follow in the next sections exploits the above two classes of WI for particular choices of the operators $\mathcal{O_P}$ and $\mathcal{O_S}$. 
In particular, the choices $\mathcal{O_P}=P^a$ in~\eqref{wigen} and $\mathcal{O_S}=S^a$ in~\eqref{wigenvec} will lead to identities between different combinations of quark condensates and susceptibilities, whereas  choosing $\mathcal{O_P}=P^aS^b$ in~\eqref{wigen} will allow one to relate differences of correlators of degenerated partners with three-point functions related to physical interaction processes. 

Those identities will involve $P$ and $S$ correlators and their corresponding susceptibilities, defined as:
\begin{eqnarray}\label{eq:chip-def}
  \chi_P^{ab}(T)&=&\int_T dx  \langle \mathcal{T} P^a (x) P^b (0)\rangle,\\
  \tilde\chi_S^{ab}(T)&=&\int_T dx  \left[\langle \mathcal{T} S^a (x) S^b (0)\rangle-\mean{S^a}\mean{S^b}\right],\label{eq:chis-def}
\end{eqnarray}
where $\int_T dx\equiv \intT$ at finite temperature $T=1/\beta$ and $\langle \mathcal{T}\cdots\rangle$ denotes the time-ordered vacuum expectation value in Minkowski space-time, 
which corresponds to a thermal average in Euclidean space-time.   Note that in the scalar case, the subtraction of the quark-bilinear expectation values $\mean{S^a}\mean{S^b}$, 
which are non-zero for $a,b=0,8$, allows one to express the susceptibilities $\tilde\chi_S^{ab}(T)$ as mass derivatives of the corresponding quark condensate, as used customarily within the ChPT framework~\cite{GomezNicola:2010tb,Nicola:2011gq,GomezNicola:2012uc} and also  to analyze the critical behavior~\cite{Smilga:1995qf}.  However, the study of partner degeneration in the lattice is formally investigated through the analysis of unsubtracted scalar susceptibilities~\cite{Buchoff:2013nra}.  In the following we will denote by $\tilde\chi_S$ the subtracted susceptibilities and by $\chi_S$ the unsubtracted ones.

\section{Identities involving susceptibilities and quark condensates}\label{sec:twoWI}

\subsection{$I=0,1$ sector: partners and chiral pattern}
\label{sec:I01}

Let us first consider quark bilinears with $I=0,1$ in the pseudoscalar and scalar sectors. Following the notation considered in the lattice~\cite{Buchoff:2013nra, Cossu:2013uua, Brandt:2016daq},
for the $I=1$ channel we define:
\begin{align}
\pi^a=i\bar\psi_l\gamma_5\tau^a\psi_l=P^a (a=1,2,3),\quad 
\delta^a=\bar\psi_l \tau^a \psi_l=S^a (a=1,2,3),
\end{align}
with  $\psi_l$ the light quark doublet with components $\psi_{u,d}$. The above light states correspond physically, as long as their quark model assignment is concerned,  to the pion and to the $a_0(980)$ resonance. 

For $I=0$, we consider the pseudoscalar $\eta^{0,8}=P^{0,8}$ and the scalars $\sigma^{0,8}=S^{0,8}$, 
as well as their combinations $\eta_{l,s}$ and $\sigma_{l,s}$, which form the basis of states:
\begin{align}\label{eq:lsto08}
\eta_l=i\bar\psi_l \gamma_5 \psi_l=\frac{1}{\sqrt{3}} \left(\sqrt{2}P^0+P^8\right),\quad 
\sigma_l=\bar\psi_l \psi_l=\frac{1}{\sqrt{3}} \left(\sqrt{2}S^0+S^8\right),\\
\eta_s=i\bar s \gamma_5 s=\frac{\sqrt{3}}{3}\left(\frac{1}{\sqrt{2}}P^0-P^8\right),\quad 
\sigma_s=\bar s  s=\frac{1}{\sqrt{3}}\left(\frac{1}{\sqrt{2}}S^0-S^8\right).\nonumber
\end{align}

Note that the $\eta_l$ and $\eta_s$ (or $\eta_{0,8}$) mix to give the physical $\eta$ and $\eta'$. 
In the same way, the mixing of the $\sigma_l$ and $\sigma_s$ (or $\sigma_{0,8}$) generates the $f_0(500)$ and $f_0(980)$ resonances. 
We remark that $\eta_l$ coincides with the physical $\eta$ state in the so-called $\eta-\eta'$ ideal mixing angle $\theta^{id}=-\arcsin(\sqrt{2/3})$, 
which is achieved when the anomalous contribution from the operator $A(x)$ in~\eqref{anomalyeq} vanishes.
This limit is reached for $N_c\rightarrow\infty$ or when the $U(1)_A$ symmetry is restored and it will play an important role in our discussion below. 

The correlators of the above bilinears, which enter in  the susceptibilities in~\eqref{eq:chis-def}, are defined as:
\begin{align} 
\mean{\mathcal{T} \pi^a(x)\pi^b(0)}=&\delta^{ab}P_{\pi\pi} (x),&\mean{\mathcal{T} \delta^a(x)\delta^b(0)}=&\delta^{ab}S_{\delta\delta} (x),\nonumber\\
\mean{\mathcal{T} \eta_l(x)\eta_l(0)}=&P_{ll} (x),&\mean{\mathcal{T} \sigma_l(x)\sigma_l(0)}=&S_{ll} (x),\nonumber\\
\mean{\mathcal{T} \eta_l(x)\eta_s(0)}=&P_{ls} (x),&\mean{\mathcal{T} \sigma_l(x)\sigma_s(0)}=&S_{ls} (x),\nonumber\\
\mean{\mathcal{T} \eta_s(x)\eta_s(0)}=&P_{ss} (x),&\mean{\mathcal{T} \sigma_s(x)\sigma_s(0)}=&S_{ss} (x).\nonumber
\end{align}
They form a basis of 8 correlators involved in this sector, where from~\eqref{eq:lsto08} the light $X_{ll}$, strange $X_{ss}$ and crossed correlators $X_{ls}$
can be expressed in terms of  $X_{00}=\mean{\mathcal{T} X_0(x)X_0(0)}$, $X_{88}=\mean{\mathcal{T} X_8(x)X_8(0)}$ and $X_{08}=\mean{\mathcal{T} X_0(x)X_8(0)}$, with $X=S,P$:  
\begin{align}
X_{ss}=&\frac{1}{3}\left(X_{88}+\frac{1}{2}X_{00}-\sqrt{2}X_{08}\right),\\ 
X_{ll}=&\frac{1}{3}\left(X_{88}+2X_{00}+2\sqrt{2}X_{08}\right),\\ 
X_{ls}=&\frac{1}{3}\left(-X_{88}+X_{00}-\frac{1}{\sqrt{2}}X_{08}\right).
\label{Xls}
\end{align}

Recall that the crossed $X_{08}$ and $X_{ls}$ correlators are in general nonzero due to mixing. Let us  give also here, for completeness, the variation of the $X^{0,8}$ bilinears under pure $SU_A(2)$ transformations, which will be of use later:
\begin{align}
\delta P^8(y)/\delta\alpha_A^a(x)&= -\sqrt{1/3}\,\delta(x-y)\delta^a(x),\nonumber\\  
\delta P^0(y)/\delta\alpha_A^a(x)&=-\sqrt{2/3}\,\delta(x-y)\delta^a(x),\nonumber\\ 
\delta S^8(y)/\delta\alpha_A^a(x)&=\sqrt{1/3}\,\delta(x-y)\pi^a(x),\nonumber\\  
\delta S^0(y)/\delta\alpha_A^a(x)&=\sqrt{2/3}\,\delta(x-y)\pi^a(x),
\label{I0su2tr}
\end{align}
with $a=1,2,3$. 

In particular, chiral axial transformations mix $\pi-\sigma_l$ and $\delta-\eta_l$ states, namely 
\begin{align}
&\delta \pi^a (y)/\delta\alpha_A^b(x)=-\delta_{ab}\delta (x-y)\sigma_l(x),&&\delta\sigma_l(y)/\delta\alpha_A^b(x)=\delta(x-y)\pi^b(x),\nonumber\\  
&\delta \delta^a (y)/\delta\alpha_A^b(x)=\delta_{ab}\delta (x-y)\eta_l(x),&&\delta\eta_l(y)/\delta\alpha_A^b(x)=-\delta(x-y)\delta^b(x)\quad\textrm{with} \ a,b=1,2,3. 
\label{chitransbi}
\end{align}

The above transformations imply then a formal degeneration of the bilinears $\pi/\sigma$ and $\eta_l/\delta$  if the chiral symmetry $SU(2)_V\times SU(2)_A\sim O(4)$ was completely restored. In other words, these bilinears would become chiral partners. In addition, $\eta_s$ and $\sigma_s$ fields are invariant under $SU_A(2)$,  as one can see from their definition~\eqref{eq:lsto08} and the transformations of the octet and singlet fields~\eqref{I0su2tr}.  
In this way, $P_{ls}$ and $S_{ls}$ transform into $\mean{\delta \eta_s}$ and $\mean{\pi\sigma_s}$ respectively, which should vanish by parity conservation. More details about particular choices of chiral rotations that implement these transformations are given in~\cite{GomezNicola:2017bhm}. 

We will use the symbol $\eqchiral$ to denote the above chiral partner equivalence. As commented in the introduction, this would actually be an exact equivalence only for two massless flavors at the phase transition.
For $N_f=2+1$ flavors and physical masses, it would become  approximate  near the crossover transition, although the equivalence is expected to be more accurate as the light chiral limit $\hat m\rightarrow 0^+$ is approached. 
Summarizing, at {\em exact} chiral restoration one has:
\begin{equation}
P_{\pi\pi}\eqchiral S_{ll}, \quad \quad P_{ll}\eqchiral S_{\delta\delta}, \quad\quad P_{ls}\eqchiral 0, \quad \quad S_{ls}\eqchiral 0, 
\label{o4deg}
\end{equation}
and so on for their corresponding susceptibilities. Therefore, the full $O(4)$ nonet partner-degeneration picture  given by the four conditions in~\eqref{o4deg},  leave four independent not degenerated correlators (or susceptibilities) in this pattern, namely  $P_{\pi\pi}$, $P_{ll}$, $P_{ss}$ and $S_{ss}$.

On the other hand, under octet and singlet axial rotations, i.e., $\alpha_A^{0,8}\neq 0$, $I=0$ states transform as:
\begin{align}
\delta P^8(y)/\delta\alpha_A^0(x)&=-\sqrt{2/3} \delta(x-y) S^8(x),&\delta P^8(y)/\delta\alpha_A^8(x)&=-\sqrt{1/3} \delta(x-y)\left(\sqrt 2 S_0- S^8(x)\right), \nonumber\\ 
\delta P^0(y)/\delta\alpha_A^0(x)&=-\sqrt{2/3} \delta(x-y) S^0(x),&\delta P^0(y)/\delta\alpha_A^8(x)&=-\sqrt{2/3} \delta(x-y) S^8(x), \nonumber\\ 
\delta S^8(y)/\delta\alpha_A^0(x)&=\sqrt{2/3} \delta(x-y) P^8(x), &\delta S^8(y)/\delta\alpha_A^8(x)&=\sqrt{1/3} \delta(x-y)\left(\sqrt 2 P_0- P^8(x)\right),  \nonumber\\ 
\delta S^0(y)/\delta\alpha_A^0(x)&=\sqrt{2/3} \delta(x-y) P^0(x), &\delta S^0(y)/\delta\alpha_A^8(x)&=\sqrt{2/3} \delta(x-y) P^8(x), 
\label{I0diagrot}
\end{align}
which allow one to mix $\pi-\delta$ and $\sigma-\eta$ states:
\begin{align} 
&\delta\pi^{a}(y)/\delta\alpha_A(x)=-\delta(x-y)\delta^a(x),&&\delta\delta^{a}(y)/\delta\alpha_A(x)=\delta(x-y)\pi^a (x),\nonumber\\ 
&\delta\sigma_l(y)/\delta\alpha_A(x)=\delta(x-y)\eta_l (x),&&\delta\eta_l(y)/\delta\alpha_A(x)=-\delta(x-y)\sigma_l (x),\quad\textrm{with}\ \alpha_A=\sqrt{1/3}\alpha_A^8+\sqrt{2/3}\alpha_A^0. 
\label{diagtrans}
\end{align}

Therefore, $\pi-\delta$ and $\sigma-\eta$  would become degenerate partners if the $U(1)_A$ symmetry was restored. Similarly,  in a fully restored $U(1)_A$ scenario, the $U(1)_A$ rotations in~\eqref{I0diagrot} and~\eqref{diagtrans} allow one to degenerate all pseudoscalar correlators into their scalar partners~\cite{GomezNicola:2017bhm}. As explained in the introduction, such restoration is only asymptotic and in general is not fully achieved in a physical $N_f=2+1$ scenario.  Nevertheless,  here we are concerned with $U(1)_A$ restoration understood as approximate partner degeneration and in that sense, we will use the symbol $\equa$. 

Thus, under $U(1)_A$ restoration the following relations hold:
\begin{equation}
P_{\pi\pi}\equa S_{\delta\delta}, \quad \quad P_{ll}\equa S_{ll}, \quad\quad P_{ss}\equa S_{ss}, \quad\quad P_{ls}\equa S_{ls},
\label{ua1deg}
\end{equation}
which leaves again four independent correlators, for instance $P_{\pi\pi}$, $P_{ll}$, $P_{ss}$, $P_{ls}$  or their corresponding scalar partners. 

Therefore, if $U(1)_A$ restoration is effective at the chiral transition, i.e., if  $O(4)\times U(1)_A$ is the restoration pattern, the four states $\pi-\delta-\sigma_l-\eta_l$ would degenerate at the transition.
Thus, the $O(4)$ and $U(1)_A$ partner equivalences in~\eqref{o4deg} and~\eqref{ua1deg} combine to $P_{\pi\pi}\sim S_{\delta\delta}\sim S_{ll}\sim P_{ll}$, which are the correlators usually analyzed in lattice works.  
Moreover, the relation $P_{ss}\sim S_{ss}$ becomes an additional signal to be analyzed. Hence, since the crossed $ls$ correlators vanish~\eqref{o4deg}, there are only two independent correlators in the $O(4)\times U(1)_A$ pattern. 

The  parameter customarily used in lattice works to parameterize the $O(4)\times U(1)_A$ degeneracy is
\begin{equation}
\chi_{5,disc}(T)=\frac{1}{4}\left[\chi_P^\pi(T)-\chi_P^{ll}(T)\right],
\label{chi5def}
\end{equation}
which vanishes at $O(4)\times U(1)_A$ restoration and  is directly related to the topological susceptibility~\cite{Buchoff:2013nra}, i.e., the correlator of the anomaly operator~\eqref{anomaly} encoding the $U(1)_A$ breaking
\begin{equation}
\chi_{top}(T)\equiv -\frac{1}{36}\chi_P^{AA}(T)=-\frac{1}{36}\int_T dx  \langle \mathcal{T} A(x) A(0) \rangle.
\label{chitopdef}
\end{equation} 

Actually, as we are about to see, the connection between $\chi_{5,disc}$ and $\chi_{top}$ is a consequence of the WI analyzed here\footnote{The normalization factor in~\eqref{chitopdef} is chosen so that $\chi_{top}$ coincides with~\cite{Buchoff:2013nra}. It comes  from our normalization of $A(x)$ in~\eqref{anomaly} and our definition of Euclidean gauge fields~\cite{Nicola:2016jlj}. Note also that the definition of susceptibilities in~\cite{Buchoff:2013nra} carries a $1/2$ normalization factor with respect to our definitions~\eqref{eq:chip-def}-\eqref{eq:chis-def}.}. 
Furthermore, in a fully $SU(3)$ restored scenario not only $\alpha^0_A$ but also  $\alpha_A^{8}$ transformations would allow one to degenerate $\pi$ and $\delta$ bilinears, hence leading to the degeneration of all other members of the scalar and pseudoscalar octet. 

More precise conclusions can be drawn from the WI in~\eqref{wigen}. The simplest choice is $\mathcal{O_P}=P^a$. Taking $\mathcal{O_P}=\pi^b, \eta_l, \eta_s$ one arrives to the following WI relating pseudoscalar susceptibilities and quark condensates analyzed in~\cite{Nicola:2016jlj,GomezNicola:2017bhm}:  
\begin{align}
&\chi_P^\pi(T)=-\frac{\condl(T)}{\hat m},
\label{wichip}\\ 
&\chi_P^{ll}(T)=-\frac{\condl(T)}{\hat m}+\frac{m_s}{\sqrt{3}\hat m(\hat m-m_s)}\chi_P^{8A}(T),
\label{chipetal}\\
&\chi_P^{ss}(T)=-\frac{\conds(T)}{m_s}+\frac{\hat m}{4\sqrt{3}m_s(\hat m - m_s)}\chi_P^{8A}(T),
\label{wichipss}
\end{align}
where $\condl=\mean{\bar\psi_l \psi_l}$  and we denote $\chi_P^{ab}=\delta^{ab}\chi^\pi_P$. 
The identities~\eqref{wichip} and~\eqref{wichipss} have been recently checked in the lattice~\cite{Buchoff:2013nra}. In the case of~\eqref{wichipss} the term proportional to $\chi_P^{8A}$ can be ignored since it is suppressed by a $\hat m/m_s$ correction. Nevertheless, it would be interesting to have a lattice check of the other WI found in~\cite{Nicola:2016jlj,GomezNicola:2017bhm} as well as those from the present work that we will discuss below.  
In particular, WI involving crossed $ls$ correlators, $I=1/2$ states and three-point functions. For the WI~\eqref{chipetal}, although the crossed $\chi_P^{8A}$ correlator is not measured in the lattice either, we will show below that this identity can be indirectly examined in the lattice with currently measured observables.  

In particular, the above identities imply that $\chi_{5,disc}$ in~\eqref{chi5def} can be written as:
\begin{equation}
\chi_{5,disc}(T)=-\frac{m_s}{4\sqrt{3}\hat m(\hat m-m_s)}\chi_P^{8A}(T)=\frac{1}{\hat m ^2}\chi_{top}(T),
\label{wichipietal}
\end{equation} 
where we have used the relation between $\chi_P^{8A}$ and $\chi_P^{AA}$ derived in~\cite{Nicola:2016jlj}, i.e., from the WI in~\eqref{wigen} taking  $\mathcal{O_P}=A$, the anomaly operator in~\eqref{anomaly}.   
The relation~\eqref{wichipietal} between $\chi_{5,disc}$ and $\chi_{top}$, also mentioned in~\cite{Buchoff:2013nra}, allows one to express $\chi_{5,disc}$ as a pure anomalous contribution, confirmed by the cancellation of the quark condensate contributions in the $\chi_P^\pi-\chi_P^{ll}$ difference in~\eqref{wichip}-\eqref{chipetal}. 

Actually, since both $\chi_{5,disc}$ and $\chi_{top}$ are measured in the lattice, although with great difficulty in the case of the topological susceptibility~\cite{Buchoff:2013nra,Borsanyi:2016ksw,Petreczky:2016vrs}, checking the relation between them in~\eqref{wichipietal} is an indirect way to check the WI~\eqref{chipetal}, or, more precisely, the combination of that identity with~\eqref{wichip}  (also checked in the lattice) and the identity connecting $\chi_P^{8A}$ and $\chi_P^{AA}$ in~\cite{Nicola:2016jlj}, namely
\begin{equation}
\chi_P^{AA}=-3\sqrt{3}\frac{\hat m m_s}{m_s-\hat m} \chi_P^{8A}.
\label{wiaa}
\end{equation}

Such verification of~\eqref{wichipietal} is actually performed  in~\cite{Buchoff:2013nra} and it holds reasonably well taking into account the difficulty to measure $\chi_{top}$. 

Now, let us turn to a very interesting relation regarding the chiral pattern, already discussed in~\cite{GomezNicola:2017bhm}, which can be obtained by analyzing the mixed  $ls$ correlators in the pseudo-scalar sector. Using~\eqref{Xls} and the relations obtained in~\cite{Nicola:2016jlj} for the susceptibilities $\chi_P^{88}$, $\chi_P^{00}$ and $\chi_P^{08}$, we get 
\begin{align} \chi_P^{ls}(T)=\frac{1}{2\sqrt{3}}\frac{1}{\hat m -m_s}\chi_P^{8A}(T),
\label{wils8A} \end{align} 
which combined with~\eqref{wichipietal}, implies:
\begin{equation}
\chi_P^{ls}(T)=-2\frac{\hat m}{m_s} \chi_{5,disc}(T)=-\frac{2}{\hat m m_s}\chi_{top}(T).
\label{wils5}
\end{equation}

The importance of the above relation is that it connects a quantity vanishing at $O(4)$ degeneration, $\chi_{ls}$ according to~\eqref{o4deg}, with $\chi_{5,disc}$ and $\chi_{top}$, signaling $O(4)\times U(1)_A$ degeneration.  
Therefore,~\eqref{o4deg} and ~\eqref{wils5} imply that if $O(4)$ partners are degenerated, so there must be the $O(4)\times U(1)_A$ ones. In other words, the chiral pattern should be $O(4)\times U(1)_A$ if {\em exact} chiral symmetry holds. 
Recall that  $\chi_{ls}\eqchiral 0$ in~\eqref{o4deg} is a consequence of the $\delta-\eta_l$ $O(4)$ degeneration~\cite{GomezNicola:2017bhm}. Thus,  more precisely,
\begin{equation}
\chi_P^{ll}\eqchiral\chi_S^{\delta} \ \Rightarrow \ \chi_{5,disc} \eqchiral 0, \quad  \chi_{top}\eqchiral 0.
\label{chi5degchiral}
\end{equation}

Several additional comments are in order here: first, the previous conclusion~\eqref{chi5degchiral} is valid in the ideal chiral restoration regime, since it relies on the $O(4)$ partner degeneration on the l.h.s. 
Nevertheless, it can be understood also in a weaker sense, as a consequence only of $\delta-\eta_l$ degeneration, which might take place approximately in a crossover scenario. 

Second, although the light chiral limit $\hat m\rightarrow 0^+$ would certainly favor exact $O(4)$ degeneration at $T_c$ and hence the realization of~\eqref{chi5degchiral}, one must not be misled by the apparent vanishing of the $\chi_{5,disc}$ term in~\eqref{wils5} when $\hat m\rightarrow 0^+$ for any $T$. This is an incorrect statement, consequence of the singular behavior of $\chi_{5,disc}$ with $\hat m$. 
Namely, at $T=0$ the results in~\cite{Nicola:2016jlj}  show that  $\chi_P^{8A}$ has a finite limit for $\hat m\rightarrow 0^+$, which together with~\eqref{wichipietal} imply $\chi_{5,disc}\sim 1/{\hat m}$ and $\chi_{top}\sim {\hat m}$  away from $T_c$.  The latter behavior for $\chi_{top}$ is actually supported in the recent work~\cite{ Azcoiti:2016zbi}, where it is argued that $\chi_{top}\sim {\hat m}\condl$ in the chiral limit.  More discussion about the chiral limit of the different susceptibilities will be carried out within ChPT in Section~\ref{sec:chpt}. 

Therefore, the vanishing of $\chi_{5,disc}$ and $\chi_{top}$ in~\eqref{chi5degchiral} are true consequences of chiral restoration. 
Similar conclusion can be drawn considering other bilinear rotations.  Namely, since $A$ is invariant under a $SU_A(2)$ transformation, the rotation $\chi_P^{8A}\rightarrow \chi^{\delta A}$ suggests $\chi_P^{8A}$ to vanish at exact chiral restoration by parity. Consequently, through~\eqref{wichipietal}, $\chi_{5,disc}$ and $\chi_{top}$ should also vanish in this limit. 

The same conclusion about the vanishing of $\chi_{top}$ for any temperature above chiral restoration has been reached in~\cite{ Azcoiti:2016zbi}. The main argument in~\cite{ Azcoiti:2016zbi} relies on the identity
\begin{equation}
\chi_P^{ ll}(T)=-\frac{\condl(T)}{\hat m}-\frac{4}{\hat m^2}\chi_{top}(T),
\end{equation}
which is nothing but the combination of~\eqref{chipetal}, \eqref{wiaa} and~\eqref{chitopdef}. In turn, note that~\eqref{wichipss} gives for the pure strange contribution
\begin{equation}
\chi_P^{ ss}(T)=-\frac{\condl(T)}{m_s}-\frac{1}{m_s^2}\chi_{top}(T),
\end{equation}
which corresponds the one-flavor version of the same identity~\cite{ Azcoiti:2016zbi}.

\begin{figure}
\centerline{\includegraphics[width=9cm]{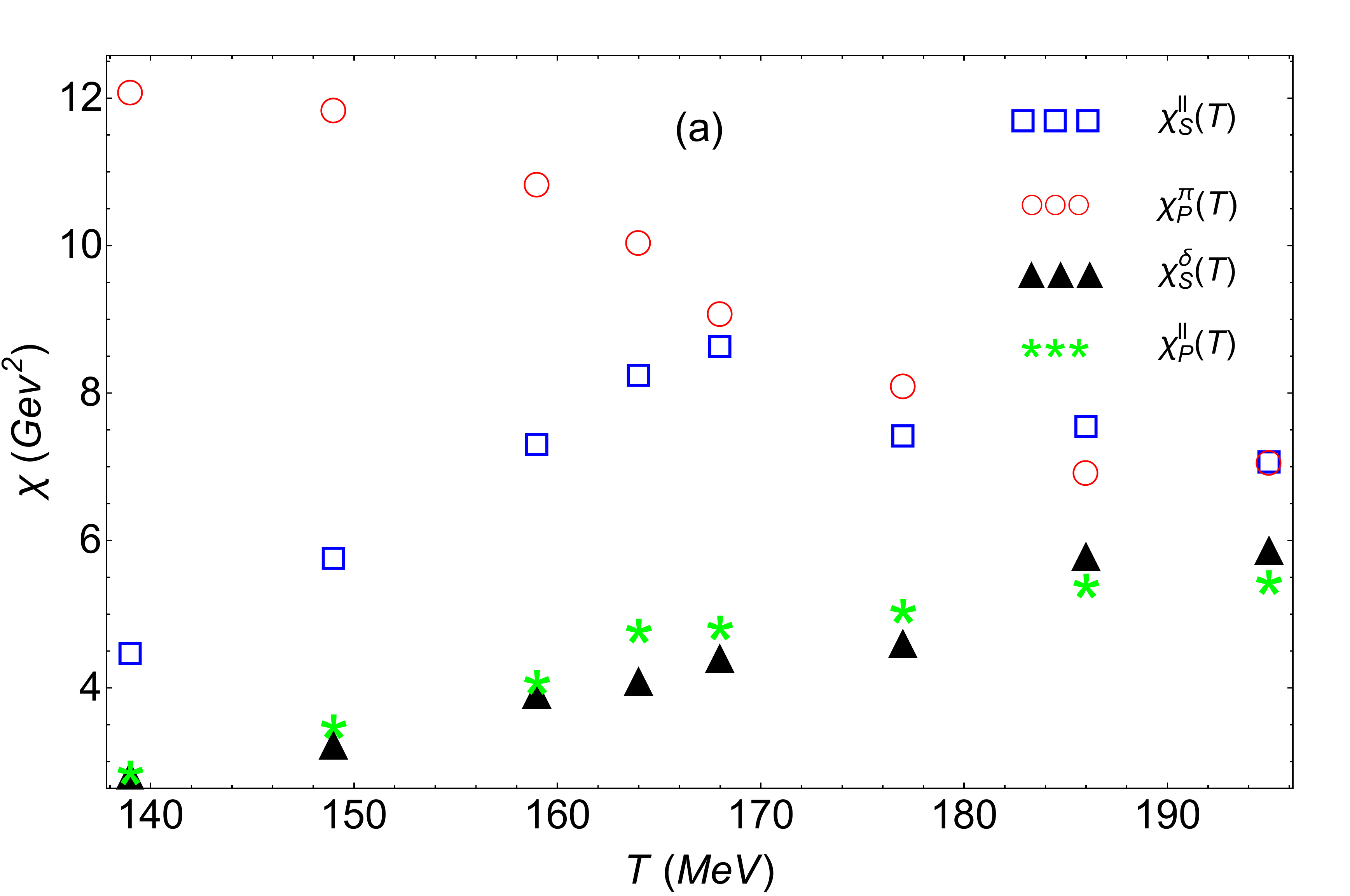}\includegraphics[width=9cm]{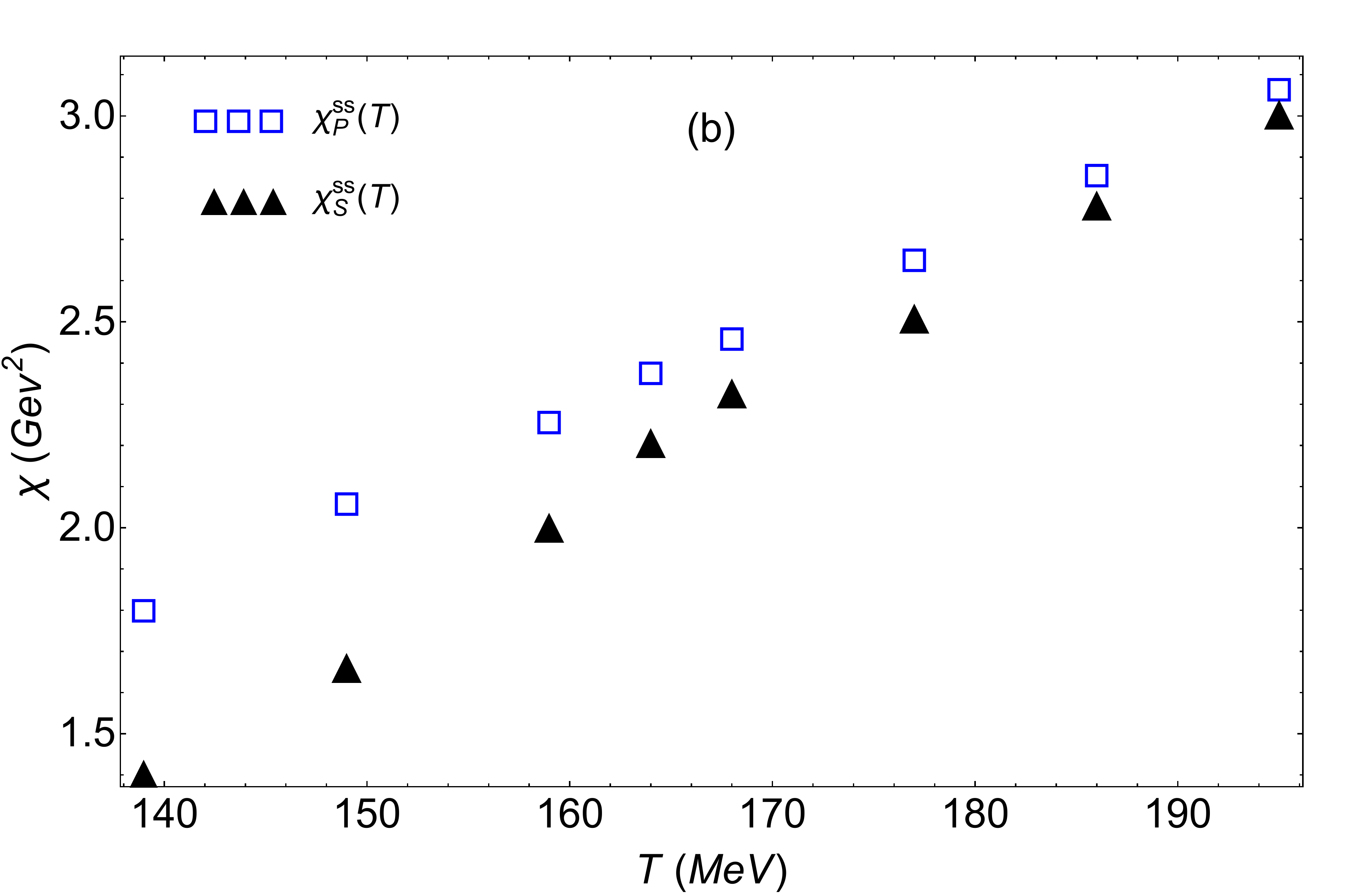}}
\caption{Different susceptibilities combinations from the lattice data in~\cite{Buchoff:2013nra} for $32^3\times 8$ lattice size. (a): The four light susceptibilities. (b): Scalar and Pseudo-scalar pure strange susceptibilities.}
\label{fig:latticenonet}
\end{figure}

Let us now comment in detail how the previous ideas are realized in present lattice simulations. As explained in the introduction, there is still some controversy regarding the chiral pattern and its nature.  In Fig.~\ref{fig:latticenonet}a we show the behavior of the four susceptibilities corresponding to the $\pi-\sigma-\delta-\eta_l$ correlators discussed above for the lattice data in~\cite{Buchoff:2013nra}.  In that work, the $O(4)$ partner degeneration corresponding to the first two equations in~\eqref{o4deg} is approximately realized at $T_c\simeq$ 160 MeV (corresponding to chiral restoration  signaled by the peak in $\chi_S^{ll}$) while the degeneration of the four correlators which would favor the $O(4)\times U(1)_A$ pattern according to~\eqref{ua1deg}  takes place asymptotically at higher temperatures.  
At this point it is worth mentioning that in a previous work~\cite{Nicola:2013vma}, $\pi-\sigma$ chiral partner degeneration in the light sector was also identified exploiting the WI~\eqref{wichip} by analyzing available lattice data for the (subtracted) quark condensate and for the scalar susceptibility.  

In addition, regarding $U(1)_A$ partner degeneration, the $ss$ correlators given by the third equation in~\eqref{ua1deg} are also compared with the lattice data of~\cite{Buchoff:2013nra} in Fig.~\ref{fig:latticenonet}b. 
We see that the degeneration of those $U(1)_A$ partners is reached also asymptotically, consistently with~\eqref{ua1deg} and Fig.~\ref{fig:latticenonet}a. 
As for the $ls$ correlators, there are no direct available data at the moment, as far as we are concerned. 

Nevertheless, as already mentioned in the Introduction, there is currently no full agreement in the lattice regarding partner degeneration and the corresponding chiral pattern.
In~\cite{Brandt:2016daq}, the difference between $\pi$ and $\delta$ screening masses are found to be compatible with zero at the chiral transition, hence pointing out to a $O(4)\times U(1)_A$ pattern even for massive light quarks. 
Since the screening masses are extracted from the two-point correlators, their degeneracy is a consequence of partner degeneration.  
In the chiral limit, the $O(4)\times U(1)_A$ pattern is also supported in the  analysis of~\cite{Cossu:2013uua}, which suggests $\pi-\delta-\sigma-\eta_l$ degeneration close to the chiral transition through the analysis of the correlators for those states in the overlap fermion lattice formulation. A recent analysis by the same group~\cite{Tomiya:2016jwr} confirms this result, showing $U(1)_A$ restoration in the chiral limit just above the transition.   
  
At this point one may wonder about the compatibility of our result~\eqref{chi5degchiral} with these lattice results.
Naively, one would conclude that we are consistent with the results in~\cite{Cossu:2013uua,Tomiya:2016jwr,Brandt:2016daq}  but not with~\cite{Buchoff:2013nra}. 
However, some considerations should be taken into account. The analysis in~\cite{Buchoff:2013nra} includes $N_f=2+1$ flavors and nearly physical light quark masses, which  may enhance $U(1)_A$ breaking effects and distort the ideal partner degeneration given in~\eqref{chi5degchiral}. Moreover, our result~\eqref{chi5degchiral} relies explicitly on $\delta-\eta_l$ degeneration at chiral restoration. However, examining in detail the numerical results in~\cite{Buchoff:2013nra}, one actually observes that the difference $\chi_P^{ll}-\chi_S^{\delta}$ is much less reduced near $T_c$ than $\chi_P^{\pi}-\chi_S^{ll}$, as it can be seen in Fig.~\ref{fig:latticenonet}a. 
In particular, from the data in Table IV in~\cite{Buchoff:2013nra}
\begin{equation} 
\left[\chi_P^{\pi}(T_c)-\chi_S^{ll}(T_c)\right]/\left[\chi_P^{\pi}(T_0)-\chi_S^{ll}(T_0)\right]\sim 0.2,\quad \left[\chi_P^{ll}(T_c)-\chi_S^{\delta}(T_c)\right]/\left[\chi_P^{ll}(T_0)-\chi_S^{\delta}(T_0)\right]\sim 8.1,\nonumber
\end{equation}
with $T_0=139$ MeV, the lowest temperature  available in~\cite{Buchoff:2013nra}. The error bars for the latter difference are also quite large, making this quantity compatible with zero for the whole temperature range considered. 
Nevertheless, the central values of $\chi_P^{ll}-\chi_S^{\delta}$ remain sizable up to the region where the $U(1)_A$ is approximately restored, i.e., where $\chi_P^{\pi}$ and $\chi_S^{\delta}$ almost degenerate. 
In this sense, the numerical results in~\cite{Buchoff:2013nra} are at odds with the expected chiral partner degeneration picture. 

\begin{figure}
\centerline{\includegraphics[width=14cm]{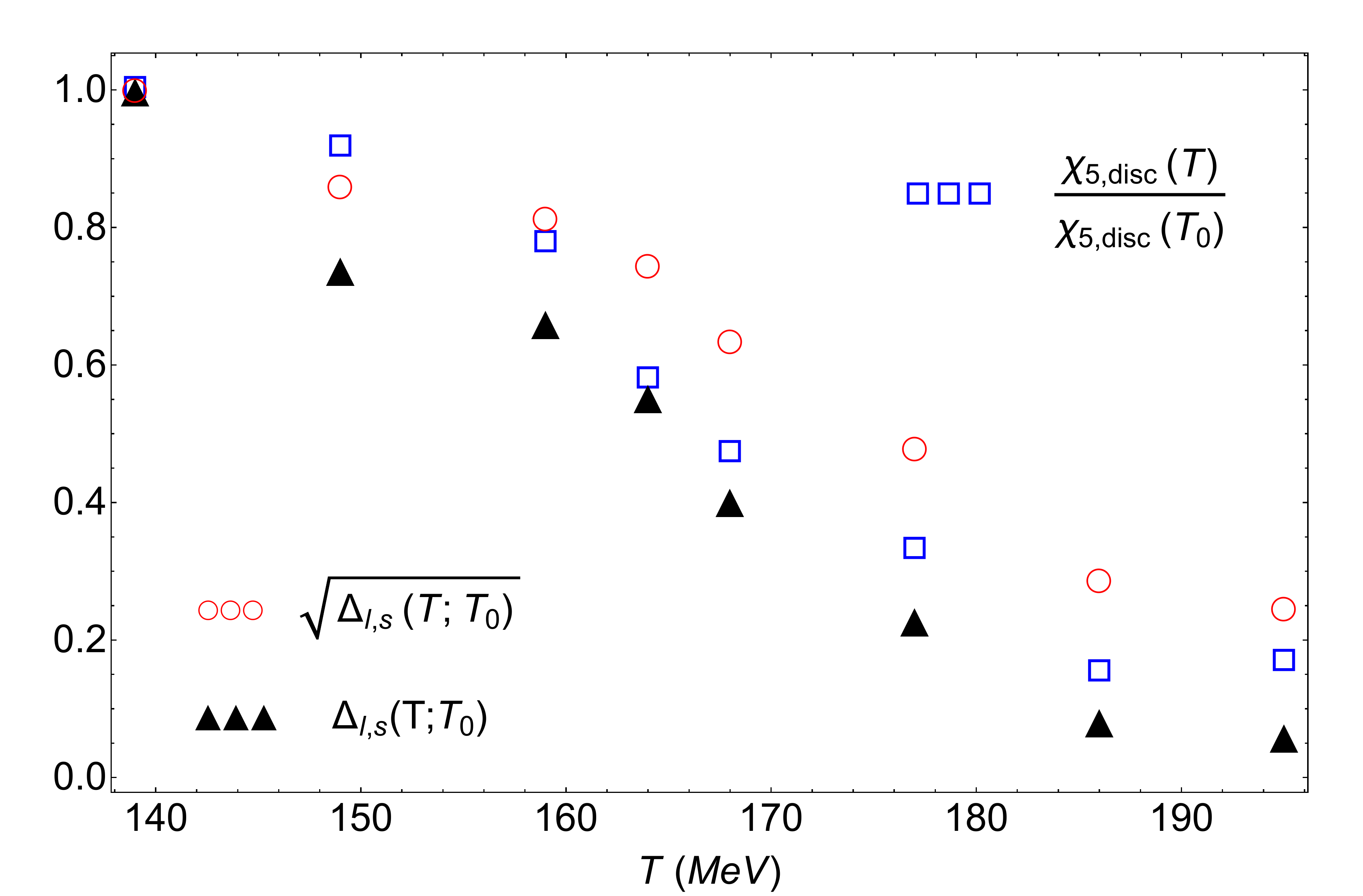}}
\caption{Comparison between the scaling of $\chi_{5,disc}$ and the subtracted condensate $\Delta_{ls}(T;T_0)$ in~\eqref{Deltals}, with respect to the reference temperature $T_0=139$ MeV. Data are taken from~\cite{Buchoff:2013nra}  for $32^3\times 8$ lattice size and $\hat m/m_s=0.088$. We include also the comparison with $\sqrt{\Delta_{l,s}}$ for the reasons explained in the main text}
\label{fig:condvschi5}
\end{figure}

On the one hand, the reasons above could explain numerically the apparent discrepancy between~\eqref{chi5degchiral} and the results in~\cite{Buchoff:2013nra}. 
On the other hand, the absence of the strange quark corrections in the $N_f=2$ lattice analysis~\cite{Cossu:2013uua,Tomiya:2016jwr,Brandt:2016daq} may explain why the $O(4)\times U(1)_A$ pattern is more clearly seen in those works, even for a finite pion mass as in~\cite{Brandt:2016daq}. A quantitative measure of the departure of the results in~\cite{Buchoff:2013nra} from the prediction~\eqref{chi5degchiral} can be achieved by comparing the temperature scaling of $\chi_{5,disc}$ with a typical chiral-restoring order parameter. Actually, as we have commented above, the analysis in~\cite{Azcoiti:2016zbi} supports $\chi_{5,dis}$ to scale with $T$ as the (subtracted) quark condensate. 
Thus, in  Fig.~\ref{fig:condvschi5} we plot $\chi_{5,disc}$ normalized to its lowest value, versus the subtracted condensate
\begin{equation}
\Delta_{ls}(T;T_0)=\frac{\condl(T)-2\frac{\hat m}{m_s}\conds (T)}{\condl(T_0)-2\frac{\hat m}{m_s}\conds (T_0)},
\label{Deltals}
\end{equation}
which is free of lattice finite-size divergences $\mean{\bar q_i q_i}\sim m_i/a^2$, with $a$ the lattice spacing, and it is one of the typical order parameters used in lattice simulations. We can see in the plot a clear correlation between the scaling of both quantities, especially near the critical region.  For comparison, we have also represented the  scaling $\sqrt{\Delta_{ls}}$, which is motivated as follows: the WI~\eqref{wichip} is compatible with the formal scaling $\pi\sim \sqrt{-\condl G_\pi^{-1}(p=0)/\hat m}$~\cite{GomezNicola:2017bhm}, which  together with~\eqref{wiPls}, to be derived in Sect.~\ref{sec:threeWI}, and~\eqref{wils5}, would lead to such square root scaling if the pion self-energy dependence with temperature is considered smooth compared to that of the quark condensate. 

\subsection{Chiral partners and mixing angles}
\label{sec:mixing}

We will explore here two interesting limits related to the mixing of the $P_0/P_8$ and $S_0/S_8$ states, namely the vanishing-mixing and ideal-mixing angles.
As we will see, these two limits are also intimately connected to the discussion of chiral partners. The mixing angle is formally defined at leading order as: 
\begin{eqnarray}
\eta=\eta_8\cos\theta_P -\eta_0\sin\theta_P,  \nonumber\\
\eta'=\eta_8\sin\theta_P  +\eta_0\cos\theta_P,
\label{mix}
\end{eqnarray}
and so on in the scalar sector with the replacements $\theta_P\rightarrow \theta_S$, $\eta\rightarrow f_0(500)$, $\eta'\rightarrow f_0(980)$. 
The mixing angle is defined to cancel the crossed $\eta\eta'$ terms in the Lagrangian, so that the correlator 
\begin{equation}
P_{\eta\eta'}=\frac{1}{2}\left(P_{88}-P_{00}\right)\sin2\theta_P  + P_{08}\cos 2\theta_P=0, 
\label{mixcond}
\end{equation}
where both, the correlators and the mixing angle, are temperature dependent. 
Let us remark that higher-order corrections introduce further mixing terms, which require additional mixing angles to be canceled. For instance at NLO in $U(3)$ ChPT two mixing angles are required~\cite{Guo:2012ym,Guo:2015xva,Gu:2018swy}. 
Nevertheless, the simplified picture above is enough for our present purposes. 

Consider first a vanishing-mixing scenario, i.e., $\theta_{P,S}=0$. 
In the pseudoscalar sector, this occurs in the pure $SU(3)$ limit, i.e., when $m_K=m_\pi$, but keeping fixed $M_0$, the anomalous contribution to the $\eta'$ mass~\cite{Guo:2012ym,Guo:2015xva,Gu:2018swy}. In that limit, $m_\eta\rightarrow m_\pi$ and $m_{\eta'}\rightarrow m_\pi+M_0$. From \eqref{mixcond},  $\theta_P\rightarrow 0$ asymptotically would imply then $P_{08}\rightarrow 0$, and so on for the scalar sector. It is important to remark that the reverse is not necessarily true. 
If  $P_{08}\rightarrow 0$ in a certain regime, we can only conclude that it implies $\theta_P\rightarrow 0$ if $P_{00}$ and $P_{88}$ remain not degenerate. According to~\eqref{Xls}, that means $P_{ls}\neq 0$. Translating these conditions to the lattice basis we conclude that in a regime of vanishing mixing angle the following conditions must hold:
\begin{align}
P_{ls}\eqnomixp P_{ll}-2P_{ss}\neqnomixp 0, \quad S_{ls}\eqnomixs S_{ll}-2S_{ss}\neqnomixs 0.
\label{nomix}
\end{align} 
In the pseudoscalar sector we can translate this result to the susceptibilities. Using~\eqref{chipetal},~\eqref{wichipss},~\eqref{wichipietal} and ~\eqref{wils5} we have
\begin{eqnarray}
2\chi_P^{ss}(T)-\chi_P^{ll}(T)+\chi_P^{ls}(T)&=&\frac{1}{\hat m}\condl (T)-\frac{2}{m_s}\conds (T)-2\frac{(\hat m - m_s)(\hat m + 2m_s)}{m_s^2}\chi_{5,disc} (T)\nonumber\\
&=&\frac{\hat m + m_s}{2m_s^2}\condl(T)-\frac{2}{m_s}\conds (T)+\frac{1}{2}\frac{(\hat m - m_s)(\hat m + 2m_s)}{m_s^2}\chi_P^{ll}(T),
\label{wipnomix}
\end{eqnarray}
where in the second line the WI~\eqref{wichip} has been used. 
This equation vanishes  in the $SU(3)$ degenerate limit, i.e., when  $m_s\to\hat m$ and $\conds\to \condl/2$. This is consistent with our previous comment since in that limit $\theta_P\to0$ and $P_{08}\to 0$. 
In addition, taking only the leading order in the  $\hat m\ll m_s$ expansion, the r.h.s. of~\eqref{wipnomix} becomes: 
\begin{equation}
\lim_{\hat m\ll m_s}\left[2\chi_P^{ss}(T)-\chi_P^{ll}(T)+\chi_P^{ls}(T)\right]=\frac{1}{2m_s}\condl (T)-\frac{2}{m_s}\conds (T)-\chi_P^{ll}(T).\label{wipnomixsmallm}
\end{equation}
Since $\condl$ is small around the chiral transition, $-\conds$ is positive and smoothly decreasing with $T$ and $\chi_P^{ll}$ is positive and increasing, 
it is plausible to expect that~\eqref{wipnomix} might be small or even vanish  near the chiral transition, although it is unclear that it should remain asymptotically small for higher temperatures. 
In addition, as we have commented above, the vanishing-mixing scenario requires $P_{ls}\neq 0$ as well, which from~\eqref{wils5} can be directly linked to $O(4)\times U(1)_A$ restoration. 
In a scenario where the chiral $O(4)$ pattern is well separated from $U(1)_A$ restoration, for instance in~\cite{Buchoff:2013nra}, it would be then  possible to find an intermediate region, roughly between chiral restoration and the $U(1)_A$ one, where the pseudoscalar mixing-angle vanishes. 

\begin{figure}
\centerline{\includegraphics[width=9cm]{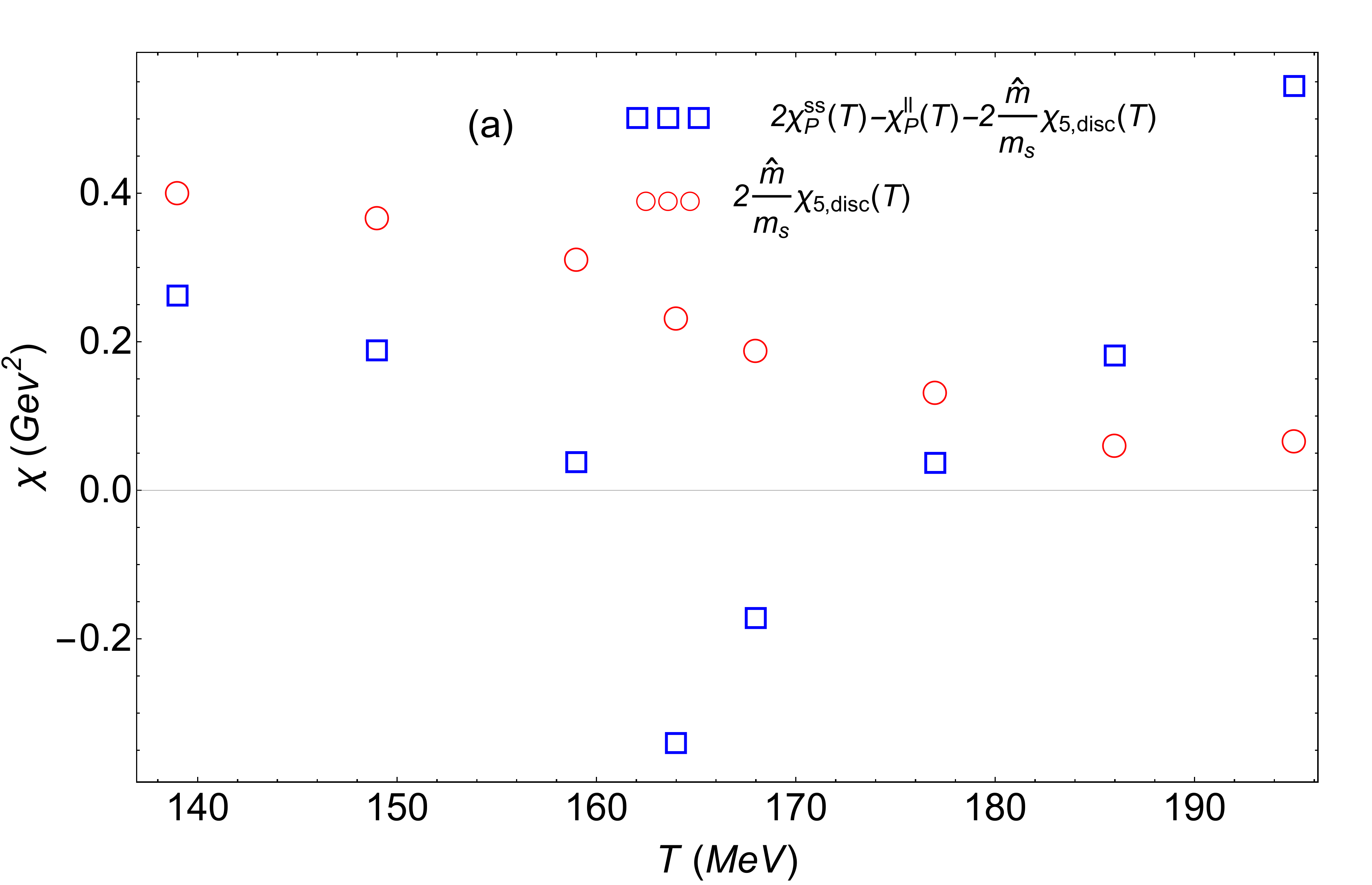}\includegraphics[width=9cm]{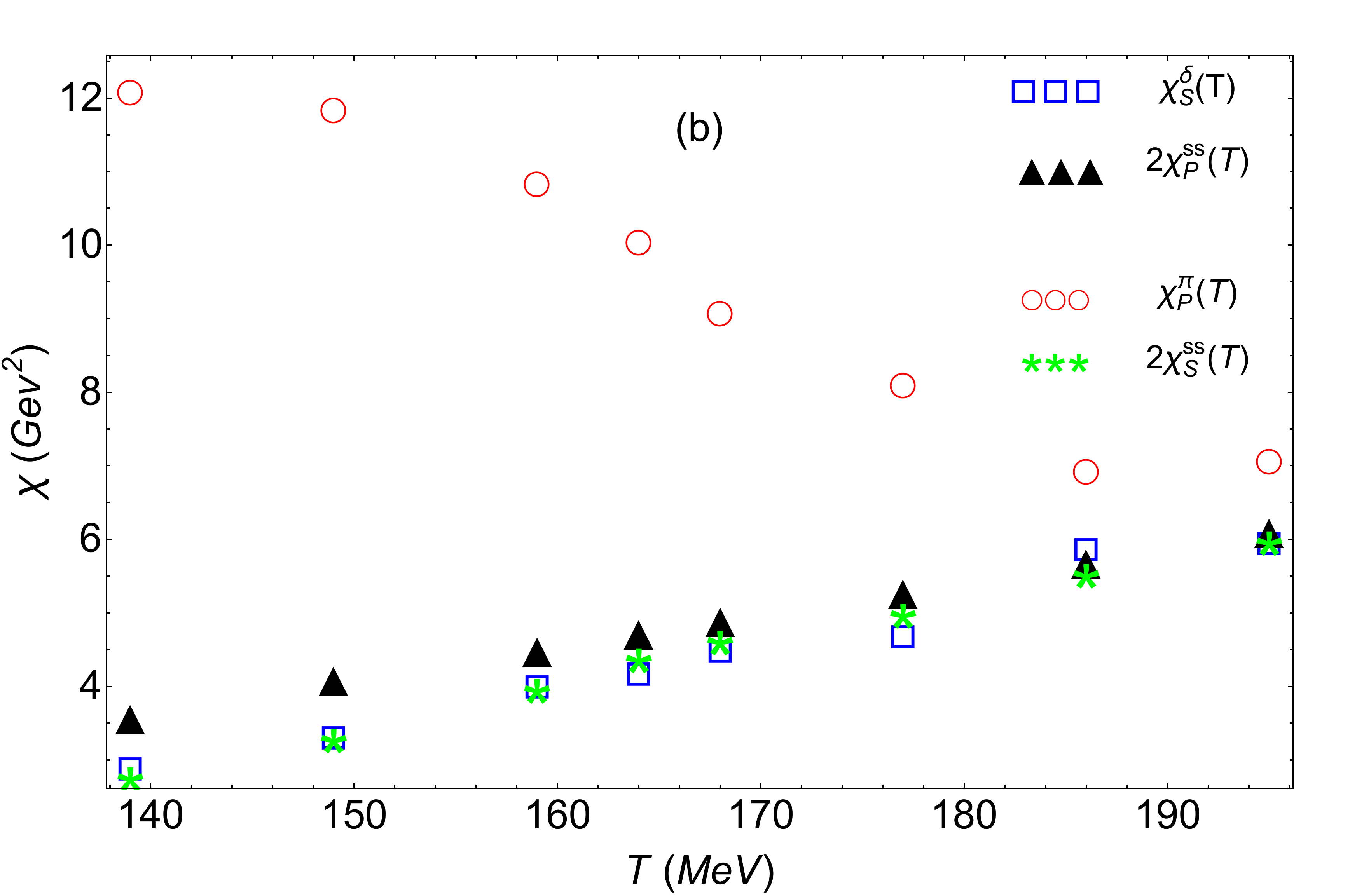}}
\caption{Different susceptibility combinations from the lattice data in~\cite{Buchoff:2013nra} for $32^3\times 8$ lattice size, related to the analysis of the $\eta-\eta'$ mixing angle.  (a): Susceptibility combination related to the vanishing of the $\eta-\eta'$ mixing angle with $\hat m/m_s=0.088$~\cite{Buchoff:2013nra}, where we also plot $-\chi_P^{ls}$ according to~\eqref{wils5}. (b): Partner degeneration in the scenario of small $X_{ls}$ and $X_{08}$ with $X=P,S$ discussed in the text, according to~\eqref{newpart}.}
\label{fig:latticenonetmix}
\end{figure}

In Fig.~\ref{fig:latticenonetmix}a we plot the susceptibility combination in the l.h.s of~\eqref{wipnomix}, signaling a vanishing of $\theta_P$, from the lattice analysis~\cite{Buchoff:2013nra}, where we have used the WI in~\eqref{wils5} for $\chi_P^{ls}$. In addition, we plot in the same figure $-\chi_{ls}=\frac{\tilde m}{m_s}\chi_{5,disc}$, which according to~\eqref{nomix} should remain nonzero to guarantee that this is a region where $\theta_P\sim 0$. 
Unfortunately, there is no way to check an analogous behavior for the scalar sector as long as $\chi_S^{ls}$ data are not provided by lattice collaborations. Consistently with our previous arguments,  we see a clear signal of the vanishing of the mixing angle, 
which happens to be very close to chiral restoration for those lattice data. Qualitatively, from the simplified $\hat m\ll m_s$ expression~\eqref{wipnomixsmallm}, the positive $-2\conds/m_s$ term dominates for low temperatures. As $T$ increases, $\chi_P^{ll}$ grows, as shown in Fig.~\ref{fig:latticenonet}a, until it compensates the strange condensate contribution. 
 The decreasing/increasing rate of $\conds$ and $\chi_P^{\eta_l}$ changes for higher temperatures, so that this susceptibility combination starts to grow again from around $T\sim$ 165 MeV, where it develops a minimum. 
Presumably, after that point the mixing angle changes from zero to the ideal one, which should be reached asymptotically at $O(4)\times U(1)_A$ restoration, consistently with the vanishing of $2\frac{\hat m}{m_s} \chi_{5,disc}(T)$, as explained below.

Consider now the ideal mixing limit $\theta=\theta^{id}=-\arcsin\left(\sqrt{2/3}\right)$, which implies that $\eta\sim \eta_l$, $\eta'\sim\sqrt{2}\eta_s$ and so on for the scalar $f_0(500)/f_0(980)$ sector. 
In a recent model analysis~\cite{Ishii:2016dln} it has been suggested that this limit can be reached from the transition temperature onwards, with a more dramatic effect for the $\eta-\eta'$ sector than for the scalar one. 
In that work, the scalar mixing remains close to ideal one for almost the entire temperature range. In the pseudoscalar sector, ideal mixing is reached when $M_0$, the anomalous contribution to the $\eta'$ mass, vanishes~\cite{Guo:2012ym,Guo:2015xva,Gu:2018swy}. In that limit, $m_\eta\rightarrow m_\pi$ and $m_{\eta'}^2\rightarrow 2m_K^2-m_\pi^2$. Thus, this limit is linked to $O(4)\times U(1)_A$ restoration, where the $\pi$ degenerates with the $\eta_l\sim\eta$, 
e.g. through the vanishing of $\chi_{5,disc}$. The strong reduction of the $\eta'$ mass observed experimentally at finite temperature~\cite{Csorgo:2009pa} supports that this limit is reached. 

From~\eqref{mixcond} and~\eqref{Xls}, we can see that $\theta_P\to\theta_P^{id}$ implies $P_{ls}\to 0$ and $S_{ls}\to 0$. 
However, as before, $P_{ls}\sim 0$ and $S_{ls}\sim 0$ are necessary but not sufficient conditions to have ideal mixing.  
Inserting $P_{ls}=0$ in ~\eqref{mixcond} leads to $\left(\sin2\theta_P-2\sqrt{2}\cos2\theta_P\right)P_{08}=0$, 
so one recovers $\theta_P=\theta^{id}$ for $\sin\theta_P<0$ and $\cos\theta_P>0$ only if $P_{08}\neq 0$. 
Therefore, in a ideal mixing regime, the following conditions must hold:
\begin{align}
P_{ll}-2P_{ss} \neqidealp P_{ls} \eqidealp 0, \quad S_{ll}-2S_{ss}  \neqideals S_{ls}\eqideals  0.
\label{idealmix}
\end{align} 

In the pseudoscalar case $P_{ls}\sim 0$ is expected at $O(4)\times U(1)_A$ restoration from~\eqref{wils5}, provided that chiral partners are ideally degenerated (formally in the chiral limit). Unlike the vanishing mixing scenario discussed above, which can be reached locally around some given temperature, ideal mixing would be reached at $O(4)\times U(1)_A$ restoration and will remain like that asymptotically. 
Thus, ideal mixing is another signal of the $O(4)\times U(1)_A$ pattern. 
In addition,~\eqref{ua1deg} implies that both the scalar and pseudoscalar mixings become ideal asymptotically. Note that as long as  $U(1)_A$ is not fully restored, $\theta_P$ and $\theta_S$ can take different values. 

Considering now the lattice results in~\cite{Buchoff:2013nra}, according to our previous argument the vanishing of $\chi_{5,disc}$ in Fig.~\ref{fig:condvschi5} signals the ideal mixing regime. 
In this regard, although one would expect to find a $\theta_P\sim 0$ region around $O(4)$ restoration, the mixing angle should turn into the ideal one as $T$ increases towards the $O(4)\times U(1)_A$ regime. 

However, it is worth mentioning that in that work the combination~\eqref{wipnomix} (blue squares in Fig.~\ref{fig:latticenonetmix}a) still remains numerically small for the temperature range explored, compared with the typical values reached by $\chi_P^{ll}$ and $\chi_P^{ss}$ in that combination (see Fig.~\ref{fig:latticenonet}a and~\ref{fig:latticenonet}b respectively). 
Thus, as $(\hat m/m_s)\chi_{5,disc}$ becomes negligible, the relation $P_{ll}  \sim 2 P_{ss}$ still holds approximately. Moreover, this condition can be combined with $P_{ll}\sim S_{\delta\delta}$, holding at $O(4)$ restoration. 
Note however that, as we discussed in Section~\ref{sec:I01},  the latter equivalence is not so accurately satisfied in~\cite{Buchoff:2013nra}. 
In conclusion, the following two additional partner degeneration conditions would be satisfied approximately in the intermediate region between $O(4)$ and $U(1)_A$ restoration:
\begin{equation}
  P_{ll}  \sim 2 P_{ss} \sim S_{\delta\delta}, \quad
  S_{ll}  \sim   2 S_{ss} \sim P_{\pi\pi}. 
\label{newpart}
\end{equation}

Near $U(1)_A$ restoration, the four correlators $2P_{ss}\sim S_{\delta\delta}\sim 2S_{ss}\sim P_{\pi\pi}$  would become degenerate. 
In Fig.~\ref{fig:latticenonetmix}b we check the degeneration~\eqref{newpart}, which holds reasonably well given the approximations considered and the lattice uncertainties.  
In fact, if the susceptibility combination in Fig.~\ref{fig:latticenonetmix}a would keep on growing for higher $T$, the degeneration in Fig.~\ref{fig:latticenonet}d would not be maintained. 

The scenario depicted in Fig.~\ref{fig:latticenonetmix} is clearly a consequence of the $O(4)$ and $U(1)_A$ neat separation in that particular lattice analysis. However, for a $O(4)\times U(1)_A$ chiral pattern, as that observed in~\cite{Cossu:2013uua,Brandt:2016daq,Tomiya:2016jwr},   there would be no room for a vanishing mixing region since $U(1)_A$ restoration is already activated around the $O(4)$ transition, where the ideal mixing would be operating.  

\subsection{Including isospin breaking: connected and disconnected scalar susceptibilities}
\label{sec:ib}

In this section, we derive additional results in the form of WI, which become useful for the discussion of the role of the connected and disconnected parts of the scalar susceptibilities regarding chiral partners and patterns. For that purpose, let us consider the general isovector isovector WI in~\eqref{wigenvec} with a scalar operator ${\cal O}^b=S^b$ satisfying
\begin{equation}
\delta{\cal O}^b(y)/\delta\alpha_V^a(x)=\delta(x-y)\epsilon^{abc}S_c,\quad\textrm{for}\quad a,b,c=1,2,3.\nonumber
\end{equation}
If we also take into account isospin breaking effects $m_u\neq m_d$ in the quark mass matrix, i.e.,
\begin{equation}
{\cal M}=\frac{1}{2\sqrt{3}}(m_u+m_d-2m_s)\lambda_8+\frac{1}{\sqrt{6}}(m_u+m_d+m_s)\lambda_0+\frac{1}{2}(m_u-m_d)\lambda_3,\nonumber
\end{equation}
the WI in~\eqref{wigenvec} becomes after integration in the Euclidean space-time 
\begin{equation}
\mean{\bar u u - \bar d d}(T)=\frac{m_d-m_u}{2} \chi_S^{\delta, ch} (T),
\label{wisusdelta}
\end{equation}
where the charged $\chi_S^{\delta, ch}=\chi_S^{11}=\chi_S^{22}$  differs in general from the neutral $\chi_S^{33}$ if $m_u\neq m_d$. 
Nevertheless, even though $\chi_S^{\delta, ch}=\chi_S^{33}=\chi_S^{\delta}$ in the isospin limit, the identity in~\eqref{wisusdelta} is nontrivial when $m_d\rightarrow m_u$, 
since $\lim_{m_d\rightarrow m_u} \frac{\mean{\bar u u - \bar d d}}{m_d-m_u}\neq 0$~\cite{Nicola:2011gq}. 
In fact, it allows one to relate the present analysis with the standard decomposition of the subtracted scalar susceptibility into its quark-diagram connected and disconnected contributions, 
which are relevant for lattice studies~\cite{Ejiri:2009ac,Buchoff:2013nra}. 
Assuming $m_u\neq m_d$ one has~\cite{Nicola:2011gq}
\begin{align}
\tilde \chi_S^{ll}&=2\tilde \chi_S^{con}+4\tilde \chi_S^{dis},\label{chidiscondecomp}\\
\tilde \chi_S^{dis}&=\tilde \chi_S^{ud},\nonumber\\
\tilde \chi_S^{con}&=\frac{1}{2}(\tilde \chi_S^{uu}+\tilde\chi_S^{dd})-\tilde\chi_S^{ud}={\partial\mean{\bar u u - \bar d d}\over\partial(m_d-m_u)}\nonumber,
\end{align}
where
\begin{equation}
\tilde\chi_S^{ij}(T)=\int_T dx  \left[\langle \mathcal{T} \bar\psi_i\psi_i (x) \bar\psi_j\psi_j (0)\rangle-\mean{\bar\psi_i\psi_i }\mean{\bar\psi_j\psi_j}\right], \quad i,j=u,d.\nonumber
\end{equation}
In this way, comparing with~\eqref{wisusdelta},  one gets 
\begin{equation}
\chi_S^{\delta}(T)=2\tilde\chi_S^{con}(T)+\Od(m_d-m_u), 
\label{chideltavschicon}
\end{equation}
consistently with recent lattice studies~\cite{Buchoff:2013nra}. The relation~\eqref{chideltavschicon} is also consistent with the $SU(3)$ ChPT isospin-breaking analysis in~\cite{Nicola:2011gq}. Our current WI derivation is completely general and then it is also valid for the $U(3)$ scenario, which will be analyzed in Section~\ref{sec:chpt}. Actually, combining~\eqref{chideltavschicon} with~\eqref{chidiscondecomp} allows one to obtain the connected and disconnected parts  from $\tilde \chi_S^{ll}$ and $\chi_S^{\delta}$, quantities which can be directly derived from the ChPT Lagrangian formulation.

In principle, the connected part of the scalar susceptibility is expected to have a softer $T$-dependence  than the disconnected one in the relevant temperature range studied here. This is observed for instance in the lattice analysis in~\cite{Buchoff:2013nra} and is confirmed in $SU(3)$ ChPT, where one finds $\tilde\chi_S^{dis}\sim T/m_\pi$ and $\tilde\chi_S^{con}\sim T^2/m_\eta^2$~\cite{Nicola:2011gq}. 
That is, the infrared (IR) $m_\pi\rightarrow 0^+$  part of $\tilde\chi_S^{ll}$ is carried only by its disconnected part, which is the perturbative counterpart of the chiral transition peak observed in the lattice for this quantity. 
Conversely, the growth of the connected piece is controlled by the heavier scale $m_\eta^2$ coming from $\pi^0\eta$ mixing and $\bar K K$ loops.  

However, it is important to remark that the above picture may change if the $U(1)_A$ symmetry is restored close to the $O(4)$ transition. First, since $\chi_S^\delta$ grows with $T$ and $\chi_P^\pi$ decreases like $\condl$ from~\eqref{wichip}, their degeneration would give rise to a maximum for $\chi_S^\delta$ at $U(1)_A$ restoration. Such possible maximum is not really seen from Fig.~\ref{fig:latticenonet}a, since higher $T$ data points in~\cite{Buchoff:2013nra} would be needed to appreciate correctly that region.   However, going back to earlier papers of the same collaboration, the observed maximum of  $\tilde\chi_S^{con}=\chi_S^\delta/2$ at around 190 MeV~\cite{Bazavov:2011nk} can be understood in this way. 

Another signal of this behavior would be a minimum of the screening mass in the $\delta$ channel (see our discussion about screening masses in Section~\ref{sec:I1/2}). 
Such minimum is clearly observed for instance in~\cite{Brandt:2016daq} and it takes place at chiral restoration. Note that the $O(4)$ and $U(1)_A$ transition almost coexist in~\cite{Brandt:2016daq}.
A minimum for the screening mass in the $\delta$-channel is also seen in an earlier work~\cite{Cheng:2010fe}. In this work, which we will refer to in Section~\ref{sec:I1/2}, 
the full $SU(3)$ degeneration is also visible at higher temperatures, where all the screening masses for different octet channels become degenerate. 

From the ChPT  point of view, the connected susceptibility peak, linked to $U(1)_A$ restoration, can be naively understood by taking the $m_\eta\rightarrow m_\pi$ limit. 
This case is reached only when the anomalous part of the $\eta'$ mass vanishes, corresponding parametrically to $U(1)_A$ restoration~\cite{Guo:2012ym,Guo:2015xva,Gu:2018swy}. 
This $m_\eta\rightarrow m_\pi$ limit generates an IR behavior for $m_\pi\rightarrow 0^+$, which will discussed in more detail in Section~\ref{sec:chpt} within the $U(3)$ ChPT framework. 

Finally, as pointed out in~\cite{Buchoff:2013nra}, from~\eqref{chi5def},~\eqref{chidiscondecomp} and~\eqref{chideltavschicon} one finds
\begin{equation}
\chi_{5,disc}(T)= \tilde\chi_S^{dis}(T)+ \frac{1}{4}\left[\chi_P^\pi(T)-\tilde\chi_S^{ll}(T)\right]+\frac{1}{4}\left[\chi_S^\delta(T)-\chi_P^{ll}(T)\right].
\label{chi5vschidisc}
\end{equation}

Since the second and third terms in the r.h.s of~\eqref{chi5vschidisc} should vanish at exact $O(4)$ restoration, then, if $U(1)_A$ is also restored $\chi_{5,disc}=0\Rightarrow \tilde\chi_S^{dis}=0$, which is an apparent contradiction with the peak for  $\tilde\chi_S^{dis}$ observed in the lattice. However, there are two possible complementary ways to address this argument: first, from the theoretical point of view, in an ideal restoration regime only the total subtracted scalar susceptibility $\tilde\chi_S^{ll}$ should be divergent at the $O(4)$ transition~\cite{Smilga:1995qf}. 
Thus, it may happen that the peak of the connected contribution at $O(4)\times U(1)_A$ transition discussed above could compensate an absent peak in the disconnected part. 
Second, in an approximate scenario where $O(4)$ and $U(1)_A$ restoration are close but still separated by a finite gap, the third term in~\eqref{chi5vschidisc} may remain small  while both $\chi_{5,disc}$ and $\tilde\chi_S^{dis}$ keep a peaking behavior scaling as $(T-T_c)^{-\gamma}/m_\pi$ in the light chiral limit, with $\gamma$ some critical exponent~\cite{Ejiri:2009ac}. 
However, at $U(1)_A$ restoration the divergent parts of $\tilde\chi_S^{dis}$ and $-\tilde\chi_S^{ll}/4$ (second term in the r.h.s of~\eqref{chi5vschidisc}) may cancel, which is compatible with a vanishing $\chi_{5,disc}$. We will actually obtain a explicit realization of this second scenario in Section~\ref{sec:chpt} in the IR limit $m_\pi\rightarrow 0^+$, where  the gap between $O(4)$ and $U(1)_A$ is also vanishing with $m_\pi$. 

\subsection{$I=1/2$: WI, partner degeneration and lattice screening masses}
\label{sec:I1/2}

Consider now transformations of the $I=1/2$ components of the octets, i.e., $P^a\equiv K^a$ and $S^a\equiv \kappa^a$ with $a=4,\dots, 7$, which correspond to the kaon (pseudoscalar) and the $\kappa$ (scalar), respectively.  
Following similar steps as before, under $SU_A(2)$ and $U(1)_A$ transformations we have: 
\begin{align}
\delta P^a (y)/\delta \alpha_A^b (x)&=-\delta(x-y)d_{abc}S^c(x),& \delta S^a (y)/\delta \alpha_A^b (x)&=\delta(x-y)d_{abc}P^c(x),\nonumber\\ 
\delta P^a (y)/\delta \alpha_A^0 (x)&=-\sqrt{2/3}\delta(x-y)S^a(x),& \delta S^a (y)/\delta \alpha_A^0(x)&=\sqrt{2/3}\delta(x-y)P^a(x),\nonumber 
\end{align}
with $a,c=4,\dots,7$ and $b=1,2,3$. 
Since there are non-vanishing $d_{abc}$ coefficients for those $a,b$ values and $c=4,\dots,7$, both $SU(2)$ and $U(1)_A$ transformations would make the $I=1/2$ $S/P$ octet partners degenerate. 
 
We will now obtain more quantitative statements studying the WI of this sector. 
On the one hand, starting with a one-point pseudoscalar operator, i.e., $\mathcal{O}^b=P^b$ with $b=4,\dots,7$, both sides of~\eqref{wigen} vanish but for $a=4,\dots,7$, for which one gets~\cite{Nicola:2016jlj} 
\begin{equation}\label{wichiK}
-(\hat m + m_s)\chi_P^K(T)=\condl (T)+2\conds (T),
\end{equation}
already obtained in~\cite{Nicola:2016jlj}.   
On the other hand, considering the isovector WI in~\eqref{wigenvec} with $\mathcal{O}^b=S^b$ ($b=4,\dots,7$) and taking into account that 
$\delta{\cal O}^b(y)/\delta\alpha_V^a(x)=\delta(x-y)f^{abc}S_c$,  we obtain: 
\begin{equation}
  \chi_S^\kappa (T)=\frac{\condl (T)-2\conds (T)}{m_s-\hat m},
  \label{wichikappa}
\end{equation}
where we have considered the isospin limit, i.e., $m_u=m_d=\hat m$ and $\mean{S^3}=\mean{\bar uu}-\mean{\bar dd}=0$, and $\chi_S^{ab}=\chi_S^\kappa \delta^{ab}$. 

This new identity~\eqref{wichikappa} has interesting consequences and provides a first hint towards the fate of $I=1/2$ partners at chiral restoration, which has not been explored yet in lattice analysis.  
Actually, the combination of~\eqref{wichiK} and~\eqref{wichikappa} gives rise to~\cite{GomezNicola:2017bhm}:
\begin{equation}
  \chi_S^\kappa (T)-\chi_P^K (T)=\frac{2}{m_s^2-\hat m^2}\left[m_s\condl (T)-2\hat m \conds (T)\right],
  \label{wichikappakaondif}
\end{equation}
which states that in the strict light chiral limit, i.e., for a second-order chiral phase transition with $\hat m=0$ and $\condl=0$ but $m_s\neq 0$ and $\conds\neq0$,  $K$ and $\kappa$ become degenerate partners. 
Moreover, in the real crossover scenario where the light quark mass and condensate are not zero,~\eqref{wichikappakaondif} provides a measure of the $I=1/2$ partner degeneracy since
\begin{equation}\label{wichikappakaondif2}
 \frac{\chi_S^\kappa (T)-\chi_P^K (T)}{\chi_S^\kappa (0)-\chi_P^K (0)}=\frac{\condl(T)-2(\hat m/m_s)\conds(T)}{\condl(0)-2(\hat m/m_s)\conds(0)}\equiv \Delta_{l,s} (T;0),
\end{equation}
with $\Delta_{ls}$ defined in~\eqref{Deltals} and, as explained above, very well determined in the lattice. 
Roughly speaking, lattice predicts $\Delta_{ls}(T;0)\sim 0.5$ at the chiral transition~\cite{Bazavov:2011nk}. Hence, according to~\eqref{wichikappakaondif2}, in the physical case $K$ and $\kappa$ would only be degenerate around 50\% of their $T=0$ value at the $O(4)$ transition. This result provides then a way to extract information on $K-\kappa$ degeneration from lattice data without measuring directly the corresponding correlators. 
It is important to remark that $K-\kappa$ correlators also degenerate at $U(1)_A$ restoration~\cite{GomezNicola:2017bhm} and then, according to the  results above, they do so at $O(4)\times U(1)_A$ restoration. A confirmation  of the previous results will be obtained also in our ChPT analysis in Section~\ref{sec:chpt}. 
 
The other important consequence of the identity~\eqref{wichikappa} is that it allows one to explain the behavior of lattice screening masses in the $\kappa$ channel, in a similar way as it was done in~\cite{Nicola:2016jlj} for the $\pi$, $K$ and $\bar s s$ ones. Actually, the only available lattice data of correlators in  this sector are the results for $K$ and $\kappa$ screening masses in~\cite{Cheng:2010fe}. 
This result shows that both screening masses degenerate beyond the chiral transition, consistently with our previous result based on~\eqref{wichikappakaondif}.  
The observed asymptotic degeneration would be a consequence of the $U(1)_A$ asymptotic restoration.

Following the analysis in~\cite{Nicola:2016jlj}, the lattice result for the $\kappa$ screening mass in~\cite{Cheng:2010fe} can also be used to check the WI in~\eqref{wichikappa}.
If we assume a smooth temperature dependence for the residue of the $\kappa$ correlator as well as for the ratio between pole and screening masses, 
we can use the WI in~\eqref{wichikappa} to obtain a prediction for the $T$ scaling of the (spatial screening) mass ratio:
\begin{equation}
\frac{M^{sc}_\kappa(T)}{M^{sc}_\kappa (0)}\sim \left[\frac{\chi_S^\kappa(0)}{\chi_S^\kappa(T)}\right]^{1/2}=\left[\frac{\condl (0)-2\mean{\bar s s}(0)}{\condl (T)-2\mean{\bar s s}(T)}\right]^{1/2},
\label{scalingkappa}
\end{equation}
since the susceptibilities correspond to zero momentum correlators and hence to inverse square masses~\cite{Nicola:2016jlj}.  
\begin{figure}
\centerline{\includegraphics[width=14cm]{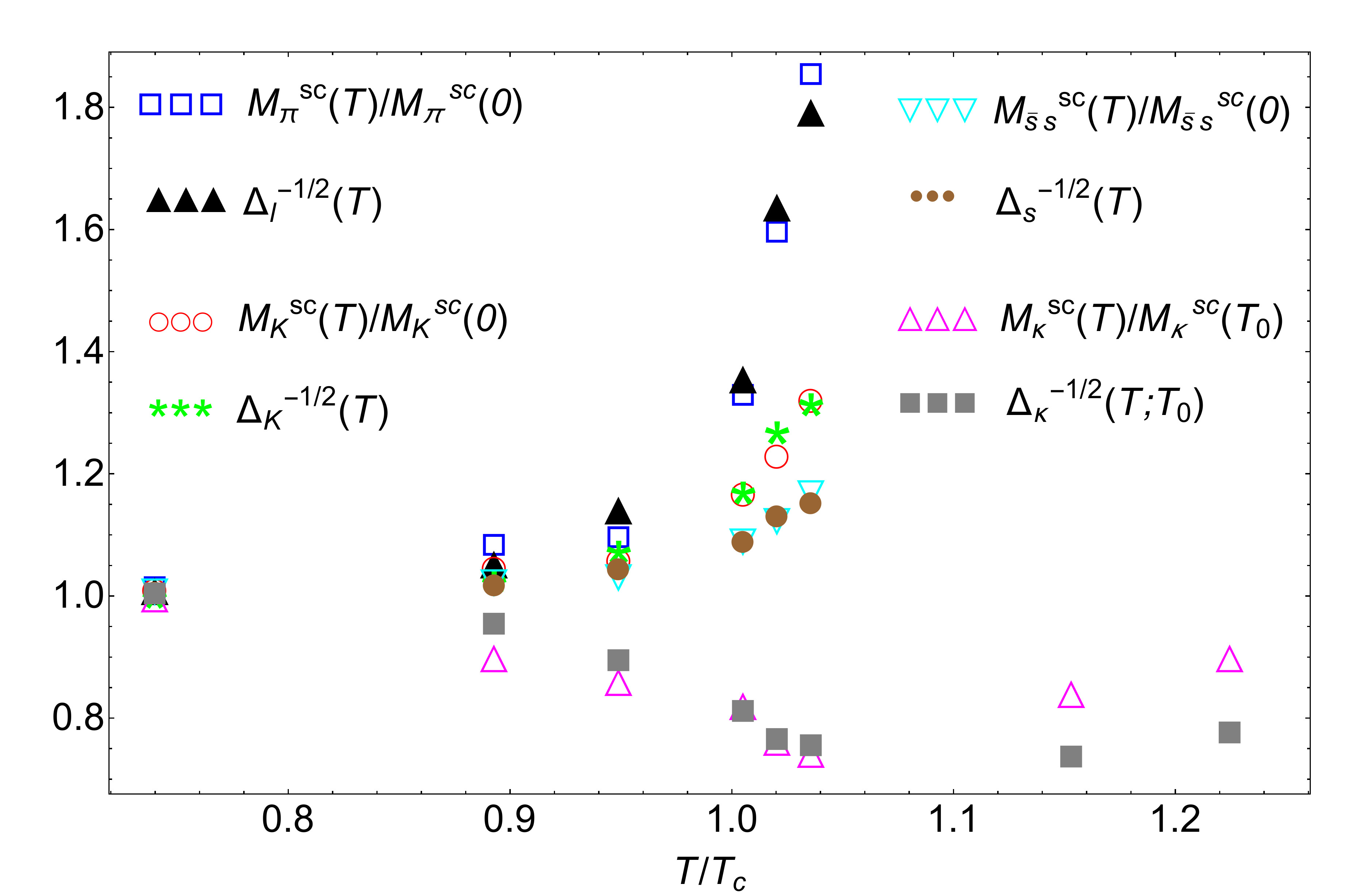}}
\caption{Comparison of pseudoscalar screening mass ratios and subtracted condensates for the four channels $\pi$, $K$, $\bar s s$ and $\kappa$ with reference values $r_1^3\condl^{ref}=0.750$, $r_1^3\conds^{ref}$=1.061.  The  lattice data are taken from 
\cite{Cheng:2010fe} (masses) and~\cite{Cheng:2007jq} (condensates) with the same lattice action and resolution, $T_0=145$ MeV, $T_c\simeq 196$ MeV and $r_1\simeq$ 0.31 fm used in~\cite{Cheng:2007jq}.}
\label{fig:corrfour}
\end{figure}

To test the scaling law in~\eqref{scalingkappa}, together with those for the $\pi$, $K$ and $\bar s s$ channels analyzed in~\cite{Nicola:2016jlj}, 
we take lattice data for screening masses and quark condensates from the same lattice group. 
As mentioned above and to the best of our knowledge, the more recent available results for screening masses in the $I=1/2$ sector are those in~\cite{Cheng:2010fe}. 
The  corresponding condensate data of the same group with the same lattice conditions ($p4$ action, $N_\tau=6$, $m_s=10\,\hat m$) are given in~\cite{Cheng:2007jq}. 
Nevertheless, as  pointed out in~\cite{Nicola:2016jlj} and in Section~\ref{sec:I01}, lattice results for quark condensates are affected by finite size divergences of the type $\mean{\bar q_i q_i}\sim m_i/a^2$.
Thus, in order to check~\eqref{scalingkappa}, we have to consider subtracted condensates free of lattice divergences. Following~\cite{Nicola:2016jlj} and~\cite{Bazavov:2011nk}, we replace $\condl (T)\rightarrow \condl (T)-\condl (0)+\condl^{ref}$ and $\conds (T)\rightarrow \conds (T)-\conds (0)+\conds^{ref}/2$, where $\condl^{ref}$ and $\conds^{ref}$ are reference values,  corresponding typically to the lattice values at $T=0$ in the chiral limit~\cite{Bazavov:2011nk}. 
We proceed as in~\cite{Nicola:2016jlj} and consider $\condl^{ref}$ and $\conds^{ref}$ as fit parameters, used to minimize the squared difference between the relative screening masses and subtracted condensates. 
We remark that we cannot just take the reference value provided in~\cite{Bazavov:2011nk} since we are taking older lattice results with very different lattice conditions.  
Thus, with only two free parameters, we can test the validity of our scaling laws based on WIs using lattice data in the four channels.
In addition, we use for the condensates the dimensionless quantity $r_1^3\mean{\bar q q}$, where $r_1\simeq$ 0.31 fm is defined in lattice analysis to set the physical scale~\cite{Cheng:2007jq,Bazavov:2011nk}. 
An important difference when including the $\kappa$ channel is that in~\cite{Cheng:2007jq} the data are not given relative to their $T=0$ value. 
Therefore, we have taken the lowest temperature point $T_0$ as the reference value for the screening masses in that channel, so that, according to~\eqref{scalingkappa}, we define:
\begin{equation}
 \Delta_\kappa (T;T_0)=\frac{\condl(T)-\condl(0)-2\left[\conds(T)-\conds(0)\right]+\condl^{ref}-\conds^{ref}}{\condl(T_0)-\condl(0)-2\left[\conds(T_0)-\conds(0)\right]+\condl^{ref}-\conds^{ref}},
\label{deltakappa}
\end{equation}
and then we should compare $M^{sc}_\kappa (T)/M^{sc}_\kappa (T_0)$ with $\Delta_\kappa(T;T_0)^{-1/2}$. 

In Fig.~\ref{fig:corrfour} we show our results for the four channels. The definitions of $\Delta_l$, $\Delta_K$ and $\Delta_s$ are given in~\cite{Nicola:2016jlj} and correspond to the subtracted condensate combinations predicted by 
the WI with respect to the $T=0$ values. It is important to point out that we have not included in the fit the points above $T_c$ in the $\kappa$ channel. 
We do not expect that the smoothness assumptions we are using to justify the scaling law can be maintained above $T_c$. 
In particular, the deviations between pole and screening masses can be sizable, as commented in~\cite{Nicola:2016jlj} and confirmed by recent model analysis~\cite{Ishii:2016dln}. 
Nevertheless, we include those points in the plot to emphasize the minimum around $T_c$ exhibited by the $\kappa$ screening mass. 
The results below $T_c$ show an excellent agreement with the predicted WI scaling, the maximum deviation being of 5.2\% (second point in the $\kappa$ channel). 
Moreover, the  reference values $\condl^{ref}$, $\conds^{ref}$ are very similar to those obtained in in~\cite{Nicola:2016jlj} for a three-channel fit. 
In addition, we remark that the scaling law in~\eqref{scalingkappa} explains qualitatively the observed minimum of $M^{sc}_\kappa$ near the transition, 
which arises from the relative behavior of (subtracted) light and strange condensates. 
Near the chiral transition the inflection point of $\condl$ signals an abrupt decreasing with respect to $\conds$, which remains smoothly decreasing.

\section{Identities relating correlator differences with three-point vertices}\label{sec:threeWI}

\subsection{$I=0,1$}

Further relations can be obtained from the axial WI in~\eqref{wigen} when two-point field operators are chosen. 
In particular, the evaluation of~\eqref{wigen} with  $\mathcal{O}^b(y)=\sigma_l(y)\pi^b(0)$ and $\mathcal{O}^b=\delta^b(y)\eta_l(0)$ 
gives rise to the identities:
\begin{eqnarray}
P_{\pi\pi}(y)-S_{ll}(y)&=&\hat{m} \int_T dx \mean{\mathcal{T} \sigma_l(y)\pi(x)\pi(0)},
\label{wipisigmaop}
\\
P_{ll}(y)-S_{\delta\delta}(y)&=&\hat{m} \int_T dx \mean{\mathcal{T} \delta(y)\pi(x)\eta_l(0)}.
\label{wideltaetaop}
\end{eqnarray}

These are particular combinations of the operators $\mathcal{O}(y)=S^{8,0}(y)\pi(0)$ and $\mathcal{O}(y)=P^{8,0}(y)\delta(0)$, which  using~\eqref{I0su2tr} yield:
\begin{widetext}
\begin{eqnarray}
P_{\pi\pi}(y)-S_{88}(y)   - \sqrt{2}S_{80}(y)&=&\sqrt{3}\hat{m} \int_T dx \mean{\mathcal{T} S^8(y)\pi(x)\pi(0)},\label{wipi8op}\\
P_{\pi\pi}(y)-S_{00}(y)   - \sqrt{\frac{1}{2}} S_{08}(y)&=&\sqrt{\frac{3}{2}}\hat{m} \int_T dx \mean{\mathcal{T} S^0(y)\pi(x)\pi(0)},\label{wipi0op}\\
P_{88}(y)    - S_{\delta\delta}(y)+\sqrt{2}P_{80}(y)&=&\sqrt{3}\hat{m} \int_T dx \mean{\mathcal{T} P^8(y)\pi(x)\delta(0)},\label{widelta8op}\\
P_{00}(y) - S_{\delta\delta}(y) +\sqrt{\frac{1}{2}}P_{80}(y)&=&\sqrt{\frac{3}{2}}\hat{m} \int_T dx \mean{\mathcal{T} P^0(y)\pi(x)\delta(0)}.\label{widelta0op}
\end{eqnarray}
\end{widetext}

Note that, due to the $\eta-\eta^\prime$ mixing, the above WIs contain the nonzero $08$ correlator, albeit it disappears in the light sector WI in~\eqref{wideltaetaop}. 
Moreover, eliminating in~\eqref{wipi8op}-\eqref{widelta0op} the $\delta\delta$ and $\pi\pi$ correlators, we get two new WIs:
\begin{widetext}
\begin{eqnarray}
P_{ls}(y)&=&
\frac{1}{3}\hat{m} \int_T dx \mean{\mathcal{T} \eta_s(y)\pi(x)\delta(0)},
\label{wiPls}\\
S_{ls} (y)&=&
-\frac{1}{3}\hat{m} \int_T dx \mean{\mathcal{T} \sigma_s(y)\pi(x)\pi(0)},
\label{wiSls}
\end{eqnarray}
\end{widetext}
which as we have seen in Sections~\ref{sec:I01} and ~\ref{sec:mixing}, play a crucial role for the discussion of the chiral pattern, partner degeneration and mixing angles.  

These identities can be translated to WIs for susceptibilities, once the integration in the $y$ variable is performed ($p=0$ in Fourier space):
\begin{align}
\chi_P^\pi-\chi_S^{ll}&=\hat m\int_T dx\, dy \mean{\mathcal{T} \sigma_l(y)\pi(x)\pi(0)},
\label{wipisigmasus}\\
\chi_P^{ll}-\chi_S^\delta&=\hat m \int_T dx\, dy \int_T dx \mean{\mathcal{T} \delta(y)\pi(x)\eta_l(0)},
\label{wietadeltasus}\\
\chi_P^{ls}&=\frac{1}{3}\hat{m} \int_T dx \, dy \mean{\mathcal{T} \eta_s(y)\pi(x)\delta(0)},
\label{wilssus}\\
\chi_S^{ls}&=-\frac{1}{3}\hat{m} \int_T dx \, dy \mean{\mathcal{T} \sigma_s(y)\pi(x)\pi(0)},
\end{align}
which can be also checked in the lattice or using different model analysis in terms of the $p=0$ three-point functions.
Note that in the $\pi-\sigma_l$ case,~\eqref{wipisigmasus} can be obtained also from  $\condl=-\hat m \chi^\pi$ using that: 
$$\int_Tdy\, dx \mean{\mathcal{T} \sigma_l(y)\pi(x)\pi(0)}=-\hat m\frac{d}{d\hat m}\chi_\pi+\int_T dx \condl \chi_\pi\nonumber.$$

The WIs in~\eqref{wipisigmaop}-\eqref{wideltaetaop} and~\eqref{wiPls}-\eqref{wiSls} parametrize the degeneration of chiral partners in terms of three-point functions. 
If $SU_A(2)$ is exactly restored, i.e., in the light chiral limit and for a vanishing light-quark condensate, the r.h.s. of these equations should vanish and hence the analysis of those three-point correlators provide alternative ways to study chiral symmetry restoration.  More precisely, according to~\eqref{wils5} and~\eqref{wiPls}: 
\begin{equation}
\chi_{5,disc}=-\frac{1}{6}m_s \int_T dx \,dy \mean{\mathcal{T} \eta_s(y)\pi(x)\delta(0)}.
\label{chi5mass}
\end{equation}

The importance of the  WIs~\eqref{wipisigmaop}-\eqref{wideltaetaop}  and~\eqref{wiPls}-\eqref{wiSls} is that they provide precise and direct information about the relevant interaction vertices and physical processes responsible for the breaking of the degeneracy in~\eqref{o4deg} in the finite mass case and $T<T_c$. In this way, the analysis of the mass and temperature dependence of the three-point functions in the r.h.s would be very relevant to analyze the evolution towards degeneration. In particular,~\eqref{wipisigmaop} and~\eqref{wideltaetaop} imply that $\pi/\sigma_l$ and $\eta_l/\delta$ partner degeneration are driven by the $\sigma\pi\pi$ and $a_0\pi\eta_l$  vertices, respectively,  whereas $a_0\pi\eta_s$ and $\sigma_s\pi\pi$ vertices enter in the crossed correlators~\eqref{wiPls}-\eqref{wiSls}.  

We could also construct WI relating three point functions in the r.h.s. of~\eqref{wipisigmaop}-\eqref{wideltaetaop} and~\eqref{wiPls}-\eqref{wiSls} with four-point pseudoscalar operators.  
This would be a much  manageable scenario within an effective theory description (like ChPT), and it would not require to introduce explicitly the $f_0(500)/(\sigma)$ degree of freedom in the Lagrangian. 
Looking in more detail at the isoscalar case, the $\sigma_l$ and $\sigma_s$ bilinears in~\eqref{wipisigmaop} couple to the scalar source $s(x)$ in the QCD Lagrangian~\cite{Gasser:1984gg}, 
which on the meson Lagrangian translates into a contribution from the $\pi\pi$, $\bar KK$ and $\eta\eta$ channels at leading order. Therefore, the r.h.s. of the identity~\eqref{wipisigmaop} is directly related to $\pi\pi\rightarrow\pi\pi$ scattering in the $I=J=0$ ($\sigma$) channel, as well as to $\bar KK\rightarrow\pi\pi$ and $\eta\eta\rightarrow\pi\pi$, where the $\sigma/f_0(500)$ resonance can also be generated.  
Thus, this identity states that the $\sigma/f_0(500)$ resonance produced in $\pi \pi$ scattering plays a fundamental role for the $O(4)$ degeneration of partners. This is fully consistent with the recent analysis in~\cite{Nicola:2013vma}, where it is shown that the critical crossover behavior of $\tilde\chi_S^{ll}$ can be achieved including the thermal pole of the $\sigma/f_0(500)$, as generated in unitarized $\pi\pi$ scattering~\cite{Dobado:2002xf}. 
Similarly, the $\delta$ bilinear translates into a contribution from the  $\pi\eta$ and $\bar K K$ channels. 
In this way, the r.h.s. of~\eqref{wideltaetaop} connects with the $a_0(980)$ resonance, which is produced in $\pi\eta\rightarrow\pi\eta$ and $\bar K K\rightarrow \pi\eta$ scattering and motivates a future finite temperature analysis of this resonance. 

Furthermore, at first glance the identities~\eqref{wipisigmaop}-\eqref{wideltaetaop} and~\eqref{wiPls}-\eqref{wiSls} 
suggest the degeneration conditions in~\eqref{o4deg} once the light chiral limit $\hat m\rightarrow 0$ is taken,
albeit this could be only possible at temperatures close to $T_c$. 
In fact, at $T=0$, $\condl$ is $\Od(1)$ in the light chiral limit and the scalar and pseudoscalar susceptibilities satisfy $\chi_P^\pi=\Od(\hat m^{-1})\gg \tilde\chi_S^{ll}=\Od(\log\hat m)$~\cite{Smilga:1995qf,GomezNicola:2012uc}, hence in contradiction with partner degenerations. 
Similarly, for the $\delta-\eta_l$ identity~\eqref{wietadeltasus}, $\chi_S^\delta=\Od(1)$ at $T=0$~\cite{Nicola:2011gq} while $\chi_P^{\eta_l}$ diverges at least as $\Od(\hat m^{-1})$~\eqref{chipetal}. 
Thus, the three-point functions in the r.h.s. of~\eqref{wipisigmaop}-\eqref{wideltaetaop} and~\eqref{wiPls}-\eqref{wiSls} should scale as $1/\hat m$ at $T=0$ in the light chiral limit.  As $T$ increases, $\chi_P^\pi$ drops proportionally to $\condl$ as given by~\eqref{wichip} while $\tilde\chi_S^{ll}$ increases. Hence, they will eventually match consistently with partner degeneration around $T_c$. 
According to~\eqref{wipisigmaop} such degeneration, expressed in term  of two-point correlators, is driven by the $\sigma\pi\pi$ vertex, which becomes the physically relevant interaction. 
The same happens in the $\delta$ channel, where $\chi_P^{ll}$ drops, hence tending to match with $\chi_S^\delta$, driven by $a_0\pi\eta$ interaction through~\eqref{wideltaetaop}. 

Further identities can be derived considering diagonal rotations $\alpha_A^0$.  On the one hand, considering $\mathcal{O}^{bc}(y)=\pi^b(y)\delta^c(0)$ and $\mathcal{O}(y)=\sigma(y)\eta_l(0)$ in~\eqref{wigen}, 
one gets for $a=0$:
\begin{widetext}
\begin{eqnarray}
P_{\pi\pi}(y)-S_{\delta\delta}(y)&=&\int_T dx \mean{\mathcal{T}\pi(y)\delta(0)\tilde\eta(x)},
\label{wipideltaop}
\\
P_{ll}(y)-S_{ll}(y)&=&\int_T dx \mean{\mathcal{T}\eta_l(y)\sigma_l(0)\tilde\eta(x)},
\label{wietasigmaop}
\end{eqnarray}
\end{widetext}
where 
\begin{equation}
\tilde\eta(x)=\hat{m}\eta_l(x)+m_s\eta_s(x)+\frac{1}{2}A(x).
\label{etatilde}
\end{equation}

On the other hand, from the transformation in~\eqref{I0su2tr}, taking the combinations $\mathcal{O}=P^{8,0}S^{8,0}$, one obtains:
\begin{widetext}
\begin{eqnarray}
\mean{\mathcal{T}P^{8,0}(y)P^{8,0}(0)}-\mean{\mathcal{T}S^{8,0}(y)S^{8,0}(0)}&=&\int_T dx \mean{\mathcal{T}P^{8,0}(y)S^{8,0}(0)\tilde\eta(x)},
\label{wi80op}
\end{eqnarray}
\end{widetext}

The identities~\eqref{wi80op} can also be combined to give for the $ls$ and $ss$ correlators:
 \begin{eqnarray}
 P_{ls}(y)-S_{ls}(y)&=&\int_T dx \mean{\mathcal{T}\eta_l(y)\sigma_s(0)\tilde\eta(x)},
\label{PlsminusSlsWI}\\
P_{ss}(y)-S_{ss}(y)&=&\int_T dx \mean{\mathcal{T}\eta_s(y)\sigma_s(0)\tilde\eta(x)}.
\label{PssminusSssWI}
 \end{eqnarray}

Like in the previous discussion, the above identities show the different vertices responsible for the symmetry breaking of the expected $U(1)_A$ degenerated patterns, i.e., $\pi-\delta$ and $\eta_{l,s}-\sigma_{l,s}$ degeneration, which are now related with additional three-point vertices.
Compared to the previous identities~\eqref{wipisigmaop}-\eqref{wiSls}, there are two new terms. Namely, one proportional to $m_s$ and an anomalous term proportional to $A(x)$ in~\eqref{etatilde}.  
The latter corresponds to the  $U(1)_A$ breaking contributions  in~\eqref{anomalyeq}.

Recall that the $A(x)$ operator couples to the  $U(1)_A$ anomalous-source $\theta(x)$, which in the meson sector and at leading order is given by $M_0^2\eta_0$, with $\eta_0$ the pseudo-scalar singlet field and $M_0^2$ a constant giving the anomalous part of the $\eta_0$ mass. All this will be discussed in detail within the $U(3)$ ChPT formalism in Section~\ref{sec:chpt}. Moreover, as discussed above, the octet $\eta_8$ and singlet $\eta_0$ fields mix to give the physical $\eta-\eta'$ states. 
In this way, the identity~\eqref{wipideltaop} can be expressed in terms of $\pi\eta(\eta^\prime)\rightarrow\pi\eta(\eta^\prime)$ and $\bar K K\rightarrow \pi\eta(\eta^\prime)$ processes in the $a_0(980)$ channel, 
whereas~\eqref{wietasigmaop},~\eqref{PlsminusSlsWI}~and~\eqref{PssminusSssWI} refer to $\eta(\eta^\prime)\eta(\eta^\prime)\rightarrow\pi\pi$, $\eta(\eta^\prime)\eta(\eta^\prime)\rightarrow \bar K K$ and $\eta(\eta^\prime)\eta(\eta^\prime)\rightarrow\eta(\eta^\prime)\eta(\eta^\prime)$ in the $\sigma$ channel. 

\subsection{$I=1/2$}

Further relations for the $K$ and $\kappa$ correlators can be obtained taking the two-point operator $\mathcal{O}^{bc}=P^b(y) S^c(0)$. 
Considering a $SU_A(2)$ transformations in~\eqref{wigen}, i.e., taking $a=1,2,3$,  one obtains for the $KK$ and $\kappa\kappa$ correlators:
\begin{widetext}
 \begin{equation}
d^{abc}\left[P_{KK}(y)-S_{\kappa\kappa}(y)\right]=\hat{m} \int_T dx \mean{\mathcal{T} K^b(y)\kappa^c(x)\pi^a(0)},\qquad (a=1,2,3, \quad b,c=4,\dots,7),
\label{wiKKappaop}
\end{equation}
\end{widetext}
where we denote $\mean{P^aP^b}=\delta^{ab}P_{KK}$ and $\mean{S^aS^b}=\delta^{ab}P_{\kappa\kappa}$ for $a,b=4,\dots,7$. 

The above identity provides information of the physical processes responsible for such degeneration. The possible values for $d_{abc}=\pm 1/2$ account for the different combinations of allowed $\kappa\rightarrow K\pi$ processes, which,  within a pure light or NGB theory, are  $K\pi\rightarrow K\pi$ and $K\eta\rightarrow K\pi$. Hence,~\eqref{wiKKappaop} highlights the relevant role of the controversial $\kappa$ resonance at finite $T$ for the chiral symmetry restoration in the $I=1/2$ channel. 
 
Finally, we will also consider the effect of $U(1)_A$ transformations in this sector. Taking $\mathcal{O}^{bc}$ as before but now with $a=0$,~\eqref{wigen} gives
 \begin{widetext}
  \begin{equation}
P_{KK}(y)-S_{\kappa\kappa}(y)=
 \int_T dx \mean{\mathcal{T}K(y)\kappa(0)\tilde\eta(x)},
\label{wiKKappaopua1}
\end{equation}
\end{widetext}
 which corresponds to $\kappa\rightarrow K\eta$ and $\kappa\rightarrow K\eta'$ decays including the anomalous contribution, or $K\eta (\eta')\rightarrow K\pi$ and $K\eta (\eta')\rightarrow K\eta (\eta')$ meson scattering processes in the $\kappa$ channel. Note that the l.h.s. of~\eqref{wiKKappaop} and~\eqref{wiKKappaopua1} are the same except for the $d_{abc}=\pm1/2$ factor, which allows one to connect the different scattering processes involved. 
 
 Thus, the vanishing of the r.h.s. of equations~\eqref{wiKKappaop} and~\eqref{wiKKappaopua1} would be consistent with the $K-\kappa$ degeneration at chiral and $U(1)_A$ transitions described in Section~\ref{sec:I1/2}. 

\section{Effective theory analysis within $U(3)$ Chiral Perturbation Theory}\label{sec:chpt}

The WI studied in this work have been derived within the QCD generating functional. Thus, in principle, they are subject to renormalization ambiguities related to the fields and vertices involved~\cite{Bochicchio:1985xa, Boucaud:2009kv}. It is therefore important to provide a specific low-energy realization of WI and the observables entering them, such  as the scalar and pseudoscalar susceptibilities that we have been analyzing in previous sections. 
We will carry out such analysis in this section, where we provide a thorough ChPT $U(3)$ analysis, hence extending  the work in~\cite{Nicola:2016jlj} to include the relevant chiral and $U(1)_A$ partners.  
As we are about to see, this study will confirm our previous findings based on WI and symmetry arguments.

The $U(3)$ ChPT formalism provides a consistent framework for calculating low-energy physical processes related to the pseudoscalar nonet. With respect to standard $SU(3)$ ChPT, where pions, kaons and the octet $\eta_8$ state are the pseudo-Goldstone bosons, it incorporates also the singlet $\eta_0$ as a ninth pseudo-Goldstone boson. 
However, due to the $U_A(1)$ anomaly, the mass of the $\eta_0$ is too heavy to be included in the standard chiral power counting in terms of meson masses, energies and temperatures.
Nevertheless, the axial anomaly vanishes in the $N_c\rightarrow\infty$ limit, in which the singlet field $\eta_0$ would become the ninth Goldstone boson in the chiral limit.
For that sake, the large $N_c$ limit framework must be considered~\cite{Witten:1979vv,HerreraSiklody:1996pm,Kaiser:2000gs}, so that the chiral counting is extended to include the $1/N_c$ counting. 
In this way, the expansion is performed in terms of a parameter $\delta$ such that $M^2, E^2, T^2,\hat m,m_s =\Od(\delta)$ and $1/N_c=\Od(\delta)$, where $M,E$ are typical meson masses and energies. In this counting,  the tree-level pion decay constant $F^2=\Od(N_c)=\Od(1/\delta)$, which hence suppresses loop diagrams. 
The counting of the different  Low-Energy Constants (LECs), according to their $\Od(N_c)$ trace structure, is given in detail in~\cite{HerreraSiklody:1996pm,Guo:2012ym,Guo:2015xva,Gu:2018swy}.  

In~\cite{Nicola:2016jlj}, one-point WI involving pseudoscalar susceptibilities and quark condensates were verified  within  $U(3)$ ChPT and the explicit expressions for those susceptibilities and condensates were given up to NNLO in the $\delta$ counting. Here, we will extend that work to the scalar sector, which will allow us to check our previous results based on WI for the nonet partners under $O(4)$ and $U(1)_A$ restoration. For that purpose, we consider the Lagrangian up to NNLO, namely ${\cal L}={\cal L}_{\delta^0}+{\cal L}_\delta+{\cal L}_{\delta^2}$ in the notation of~\cite{Guo:2012ym,Guo:2015xva,Gu:2018swy}. 
Besides, the $\eta-\eta'$ mixing angle has to be properly incorporated. The explicit expressions for lagrangians, self-energies and the mixing angle up to the relevant order we are considering here can be found in~\cite{Guo:2015xva}.  

Within this $U(3)$ framework and including scalar sources in the effective Lagrangian as dictated by chiral symmetry~\cite{HerreraSiklody:1996pm,Kaiser:2000gs,Guo:2012ym,Guo:2015xva,Gu:2018swy}, 
we have calculated  all  the scalar susceptibilities involved in our present analysis, namely $\tilde\chi_S^{ll}(T)$, $\tilde\chi_S^{ss}(T)$, $\chi_S^{ls}(T)$, $\chi_S^{\delta}(T)$ and $\chi_S^\kappa(T)$ up to the NNLO $\Od(\delta^0)$. 
Their explicit expressions are collected in Appendix~\ref{sec:app}. With those expressions, we have checked that the WI~\eqref{wichikappa} holds to the order considered. Therefore, together with the analysis in~\cite{Nicola:2016jlj} of the identities~\eqref{wichip}-\eqref{wichipss} and~\eqref{wichiK},  we complete the check of all the one-point WI. Recall that the LO $\Od(\delta^{-2})$ vanishes for the scalar susceptibilities (it contributes to the pseudoscalar ones).  Note also that, since we work within the Dimensional Regularization scheme, the differences  $\tilde\chi_S^{ll}-\chi_S^{ll}$ and $\tilde\chi_S^{ss} -\chi_S^{ss}$ formally vanish as $\delta^{(D)}(0)$ in the ChPT calculation.

As in the  $SU(3)$ calculation of  scalar susceptibilities~\cite{GomezNicola:2010tb,GomezNicola:2012uc}, our present calculation involves tree level terms, as well as one-loop corrections. 
Temperature effects show up on three type of topologies: 
\begin{enumerate}
\item Tadpole contributions coming from the Euclidean tree-level propagator $G_i(x=0)$, whose finite part reads
\begin{align}
\mu_i(T)=\frac{m_{0i}^2}{32\pi^2 F^2}\log\frac{m_{0i}^2}{\mu^2}+\frac{g_1(m_{0i},T)}{2F^2}, \\
g_1(M,T)=\frac{T^2}{2\pi^2}\int_{M/T}^\infty dx  \frac{\sqrt{x^2-(M/T)^2}}{e^{x}-1},
\label{mudef}
\end{align}     
where  $i=\pi,K,\eta,\eta'$, $m_{0i}$ are the tree level masses and $\mu$ is the  renormalization scale.
\item  Contributions arising from Wick contractions of two pairs of  meson fields at  different space-time points, proportional to 
\begin{align}
\int_T d^D x \ \left[G_i(x)\right]^2=-\frac{d}{d m_{0i}^2} G_i(x=0),
\label{G2}
\end{align}
whose finite part can be written in terms of
\begin{eqnarray}
\nu_i (T)&=&F^2\frac{d}{d m_{0i}^2}  \mu_i(T) = \frac{1}{32\pi^2} \left[1+\log\frac{m_{0i}^2}{\mu^2}\right]-\frac{g_2(m_{0i},T)}{2},\\
g_2(M,T)&=&\frac{1}{4\pi^2}\int_{M/T}^\infty dx  \ \frac{1}{x}\frac{1}{e^{x}-1}.
\label{nudef}
\end{eqnarray}
\item Loop contributions coming from mixed contributions of the type:
\begin{align}
\int_T d^D x \ G_i(x) G_j(x)=\frac{1}{m_j^2-m_i^2}\left[G_i(x=0)-G_j(x=0)\right],
\label{mixedterms}
\end{align}
which reduces to~\eqref{G2} for $m_i^2\rightarrow m_j^2$.
\end{enumerate}

An important consistency check of our calculation is that all the results are finite and scale independent. Together with the $\chi_P$ susceptibilities already calculated in~\cite{Nicola:2016jlj}, these results will allow us to examine how our previous results on partner degeneration are realized within ChPT. 
Although the ChPT framework is limited to a low temperature description, we are going to see that the thermal extrapolation of the ChPT curves provides useful model-independent results confirming our previous analysis for partner degeneration. In addition, this framework will allow us to examine the chiral limit consistently. 

\begin{figure}
\centerline{\includegraphics[width=8.8cm]{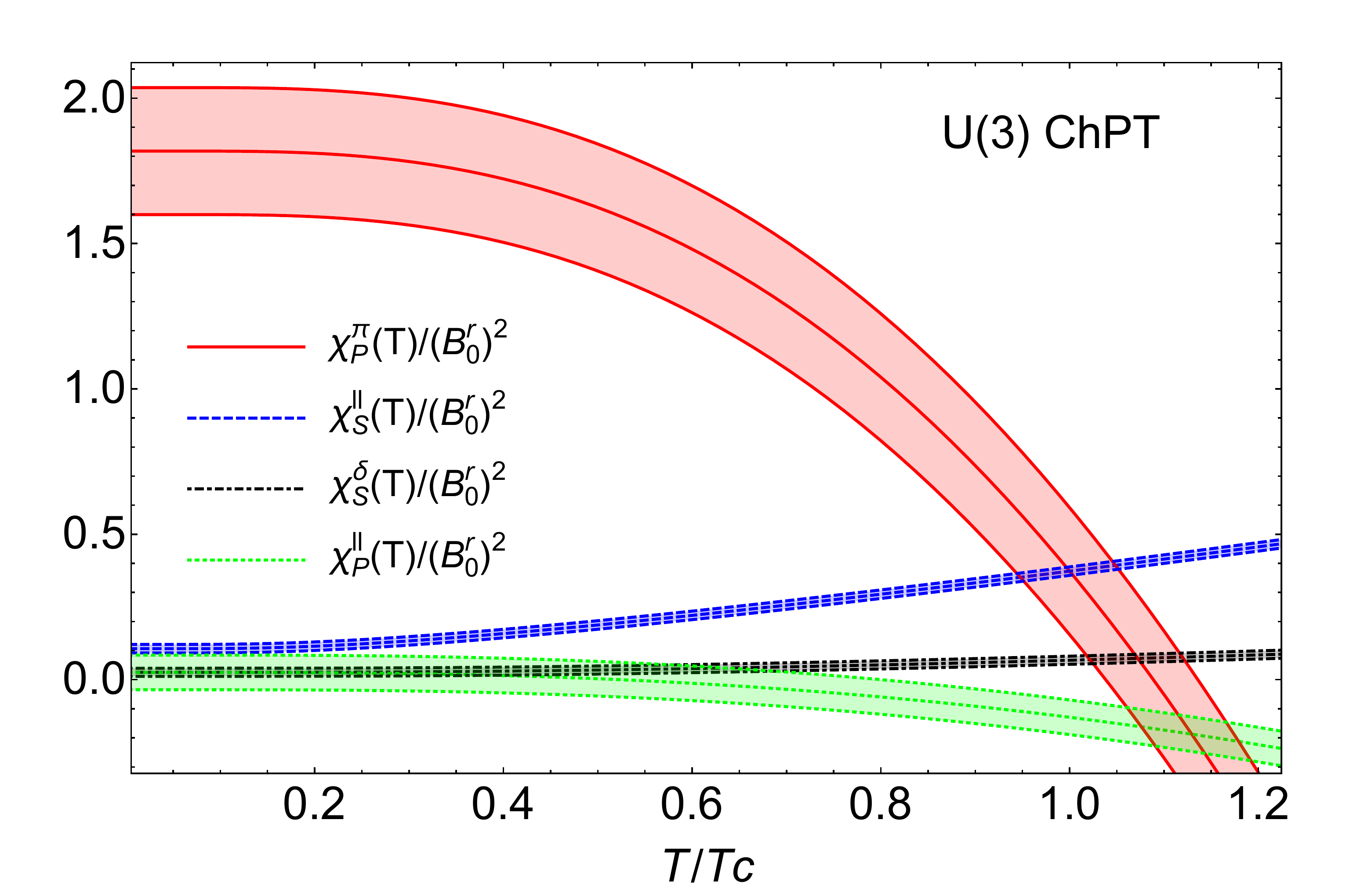}\includegraphics[width=9cm]{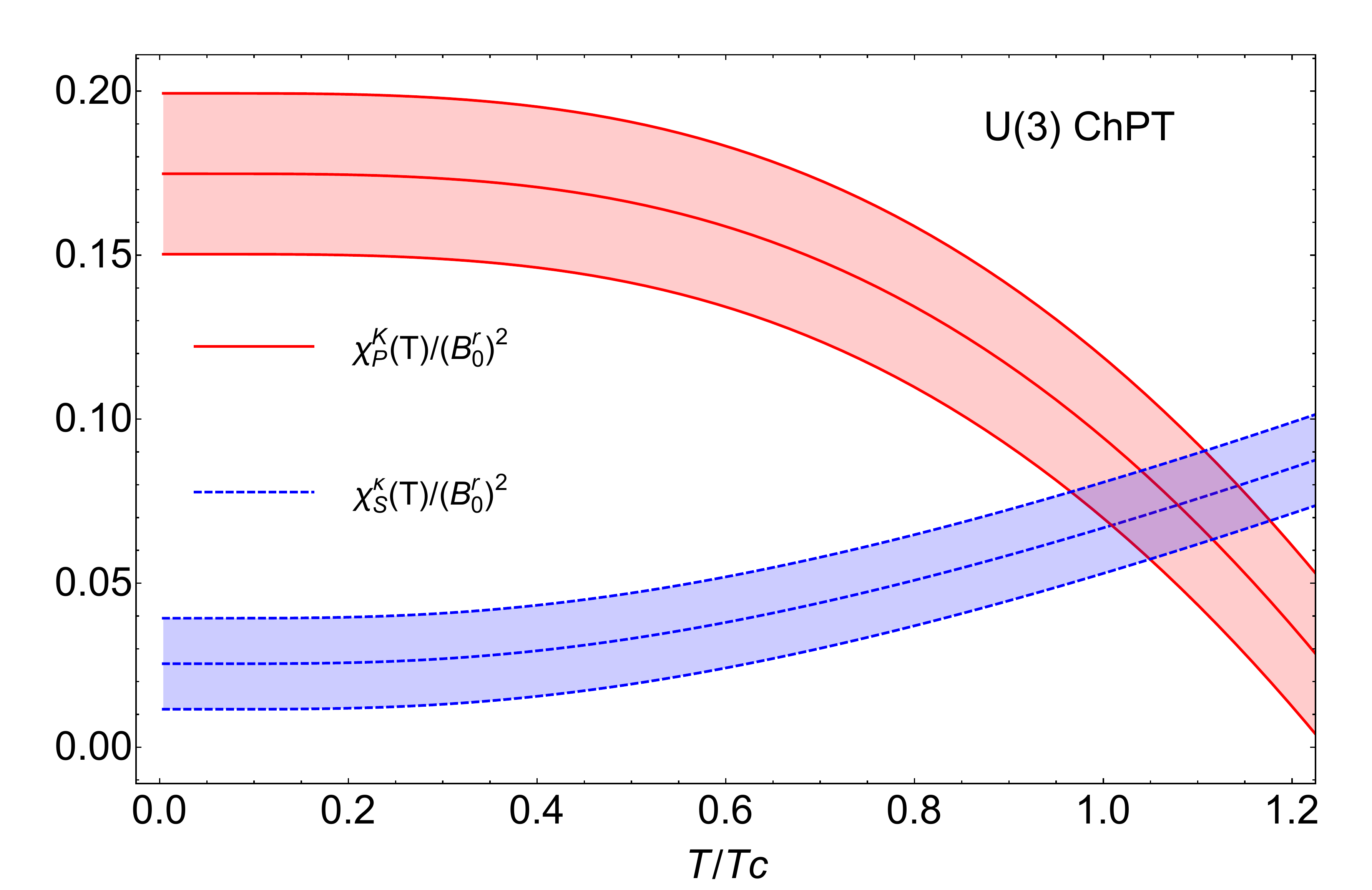}}
\caption{Susceptibilities calculated in $U(3)$ ChPT to NNLO for the physical pion mass. Left:The four susceptibilities of the $I=0,1$ sector. Right: $I=1/2$ sector susceptibilities}
\label{fig:foursuschpt}
\end{figure}

Let us start by analyzing in $U(3)$ ChPT the susceptibilities in Section~\ref{sec:I01} regarding the $O(4)$ vs $O(4)\times U(1)_A$ pattern and the corresponding partner degeneration in the $I=0,1$ sector. The results for the four susceptibilities involved are  plotted in Fig.~\ref{fig:foursuschpt} for the physical value of the pion mass. The numerical values of the LECs involved are taken from~\cite{Guo:2015xva} and the bands in the figure cover the uncertainties of those LEC quoted also in~\cite{Guo:2015xva}. 
We consider the values of the NNLOFit-B fit in~\cite{Guo:2015xva}, which is their best fit to lattice predictions of $\eta$ and $\eta'$ masses. 
All the susceptibilities are proportional to $B_0^{r\,2}=m_{0\pi}^4/(4\hat m^2)$, where, due to the presence of $\eta'$ loops, $B_0^r$ is the renormalized $U(3)$ version of the $SU(3)$ $B_0$ constant.

Let us define $T_c$ as the (pseudo-critical) $O(4)$ restoration temperature for which degeneration of the chiral partner states $\sigma/\pi$ takes place, i.e., $\chi_P^\pi (T_c)=\chi_S^{ll} (T_c)$.
Note that this temperature is more advisable than the standard definition in terms of the vanishing quark condensate, since the latter is meant to remain nonzero at the chiral transition for physical masses. 
Recall that  throughout this section, what we really mean by degeneration of partners is the matching of their corresponding susceptibilities, since ChPT is not able to reproduce neither a true degeneration, nor a crossover or a phase transition behavior.  Numerically, for the physical pion mass and for the LECs in~\cite{Guo:2015xva}, we obtain $T_c\sim 264$ MeV and $T_0\simeq 1.09$ MeV (for the central values in Fig.~\ref{fig:foursuschpt}) where $T_0$ is defined as $\condl(T_0)=0$.  
We stress that the particular numerical value for $T_c$  is not important, i.e. the ChPT expansion is limited at  low temperatures so it is not supposed to provide a quantitative description of the transition. 
Nevertheless, as we are about to see, the main qualitative features in terms of partner degeneration and the relation between different pseudo-critical temperatures obtained from the extrapolation of the ChPT results are consistent with lattice and with our previous WI analysis.

In addition, the results in Fig.~\ref{fig:foursuschpt} show  that $\chi_P^\pi$ matches $\chi_S^\delta$ above $T_c$. This crossing point can be considered as an estimate of $U(1)_A$ degeneration with a critical temperature $T_{c2}$ defined as $\chi_P^\pi (T_{c2})=\chi_S^{\delta} (T_{c2})$. Using  physical pion masses one finds $T_{c2}\simeq 1.07\,T_c$ (for the central values) i.e., quite close to $T_c$. Nevertheless, the  numerical  difference lies within the ChPT uncertainty range, as seen in the figure.  
The behavior of $\chi_P^{ll}(T)$ shown in Fig.~\ref{fig:foursuschpt} is not so reliable as the other susceptibilities. In this case the $\Od(\delta^0)$ ChPT corrections at $T=0$ turn out to be of the same order as the leading $\Od(\delta^{-1})$ ones. This effect is worsened as $T$ increases.  
Nevertheless, taking this caveat in mind, we can still see that the difference between $\chi_P^\pi (T)$ and $\chi_P^{ll}(T)$ does vanish close to (and above) $T_{c2}$. 
Once more, this value can be taken as the pseudo-critical temperature  characteristic of $O(4)\times U(1)_A$ restoration, which according to~\eqref{chi5def} we define as $\chi_{5,disc}(T_{c3})=0$.  
In the physical case depicted in Fig.~\ref{fig:foursuschpt} we get $T_{c3}\simeq 1.13 T_c$. 
As a summary, from the results plotted in Fig.~\ref{fig:foursuschpt}, we conclude that the $U(3)$ ChPT analysis yields $O(4)\times U(1)_A$ partner degeneration close and above $O(4)$. 
Recall that we may have different pseudo-critical temperatures in terms of partner degeneration, both for $O(4)$ and for $U(1)_A$ partners, in the physical mass case.  

In Fig.~\ref{fig:foursuschpt} we also show the $K$ and $\kappa$ susceptibilities for $I=1/2$. They match at $\chi_P^K(T_{c4})=\chi_S^\kappa(T_{c4})$ with $T_{c4}\simeq T_{c2}$. 
This behavior is compatible with the pattern predicted in Section~\ref{sec:I1/2}, i.e., $K-\kappa$ degeneration takes place at $U(1)_A$ restoration. Furthermore, as we will see below, this temperature approaches  $O(4)$ restoration in the chiral limit, consistently with~\eqref{wichikappakaondif}. 

\begin{figure}
\centerline{\includegraphics[width=14cm]{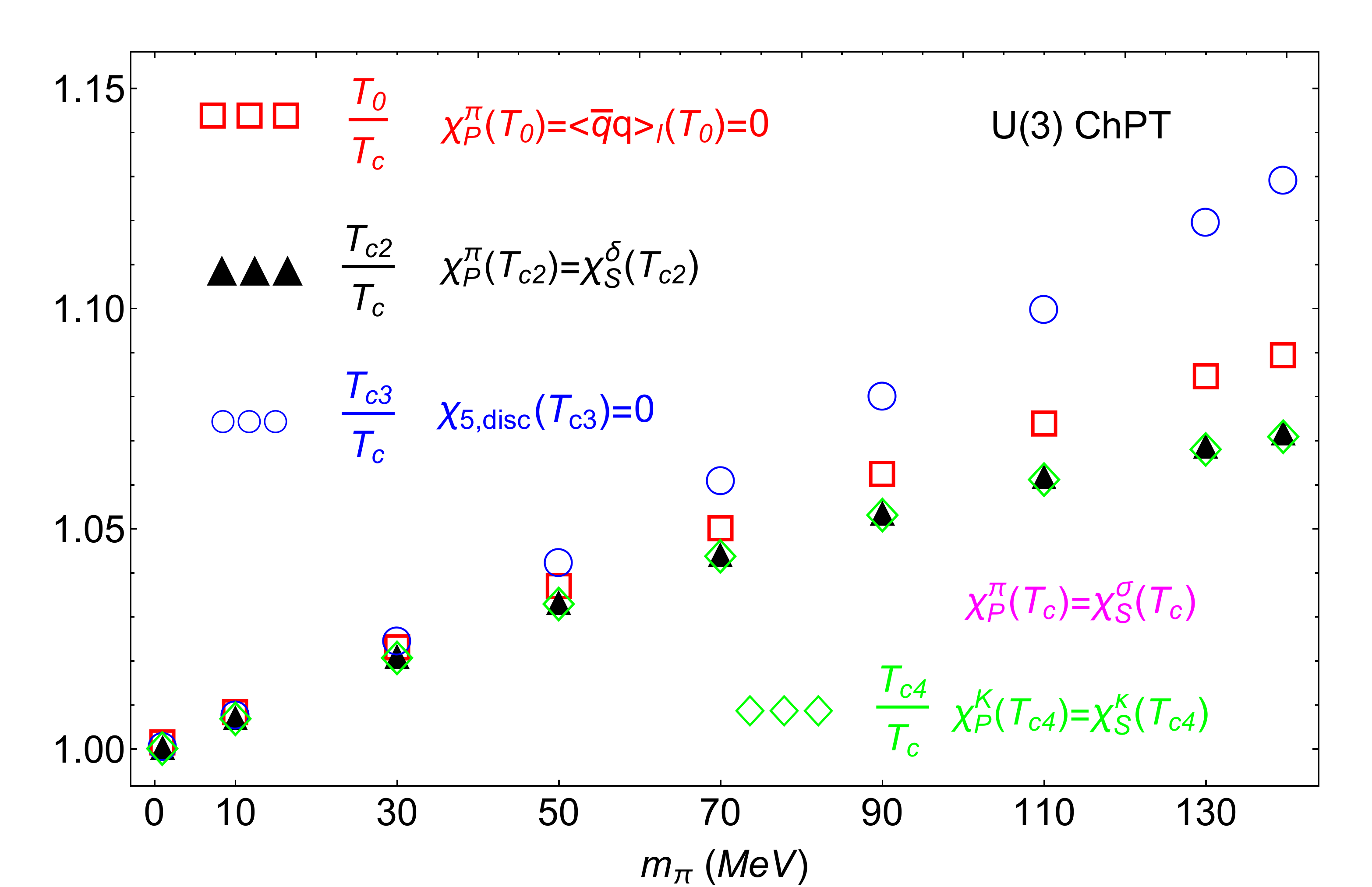}}
\caption{Different partner degeneration temperatures as the light chiral limit $m_\pi\rightarrow 0^+$ is approached.}
\label{fig:chiraltc}
\end{figure}

More revealing results are obtained from our ChPT expressions when we approach the chiral limit.  
In that regime, we would expect that the two pseudo-critical temperatures corresponding to the chiral transition, $T_0$ and $T_c$, should tend to coincide. 
In addition, from the analysis in Section~\ref{sec:I01}, we would also expect the $U(1)_A$ and $O(4)\times U(1)_A$ pseudo-critical temperatures to approach the chiral $O(4)$ ones. 
This is indeed what we obtain, as it is shown in Fig.~\ref{fig:chiraltc}, where the hierarchy $T_{c3}>T_0>T_{c2}>T_c$ is maintained as the chiral limit is approached. 

As explained above, $T_0>T_c$ is expected from chiral restoration arguments, while we expect $T_{c2}>T_c$ and $T_{c3}>T_c$  since $U(1)_A$ partners are meant to degenerate after $O(4)$ ones. 
It is also natural that $T_{c3}>T_{c2}$ since the restoration of $\chi_{5,disc}$ requires the vanishing of both $\chi_P^\pi-\chi_S^\delta$ and $\chi_S^\delta-\chi_P^{ll}$. 
In any case, from our present ChPT approach, given the decreasing behavior obtained for $\chi_P^{ll}$ in Fig.~\ref{fig:foursuschpt}, the condition $T_{c3}>T_{c2}$ clearly holds. 
Finally, there is no a priori reason on how $T_{c3}$ or $T_{c2}$ should be related to $T_0$. 

As for the $I=1/2$ $K-\kappa$ matching, we see from Fig.~\ref{fig:chiraltc} that $T_{c4}$ remains almost identical to $T_{c3}$ for all values of $m_\pi$, approaching the other restoration temperatures in the chiral limit.
This is consistent with what we expect from the WI~\eqref{wichikappakaondif}. 

Moreover, the leading order in the chiral limit for the  susceptibilities is actually quite useful for our present purposes. 
We obtain from the expressions in Appendix~\ref{sec:app}:
\begin{eqnarray}
\tilde \chi_S^{ll}(T)&=&a\left(B_0^r\right)^2 \frac{T}{m_\pi} + \Od(\log m_\pi),\nonumber\\
\chi_P^{\pi}(T)&=&\left(B_0^r\right)^2 \frac{2f_\pi^2}{m_\pi^2}\left[b\left(1-\frac{T^2}{T_{0ch}^2}\right)+a \frac{m_\pi T}{f_\pi^2}\right]+\Od(\log m_\pi),\nonumber\\
\chi_P^{ll}(T)&=&\left(B_0^r\right)^2\left(c_0-c_1\frac{T^2}{m_K^2}\right)+\Od(m_\pi),\nonumber\\
\chi_S^{\delta}(T)&=&\left(B_0^r\right)^2\left(d_0+d_1\frac{T^2}{m_K^2}\right)+\Od(m_\pi),
\label{suschiral}
\end{eqnarray}
where $a,b,c_0,c_1,d_0$ and $d_1$ are positive dimensionless constants independent of $T$ and $m_\pi$. One has $a=\frac{3}{4\pi}$, $b=\frac{\vert\condl^0\vert}{B_0^r f_\pi^2}$ and $T_{0ch}=2\sqrt{\frac{\vert\condl^0\vert}{B_0^r}}\simeq$ 238 MeV the chiral limit value for $T_0$, with $\condl^0$ the light quark condensate in the chiral limit at $T=0$. 
The analytic expressions of the other constants depend on different LECs, masses and mixing parameters and are too long to be displayed here. 
Their numerical values in the chiral limit are $c_0\simeq 0.0025$, $c_1\simeq 0.78$, $d_0\simeq0.029$ and $d_1\simeq 0.26$. The asymptotic expansions in~\eqref{suschiral} arise from:
\begin{equation}
g_1(M,T)=\frac{T^2}{12}-\frac{TM}{4\pi}+\Od(M^2\log M), \quad g_2(M,T)=\frac{T}{8\pi M}+\Od(\log M),
\label{g1g2exp}
\end{equation}
while exponentially suppressed contributions of order $\exp(-m_K/T)$ have been neglected. 

From the previous expressions and the definitions of pseudo-critical temperatures explained before, we get
\begin{eqnarray}
T_0&=&T_{0ch}+\frac{a T_{0ch}^2}{2bf_\pi^2}m_\pi+\Od(m_\pi^2\log m_\pi),\nonumber\\
T_c&=&T_0-\frac{a T_{0ch}^2}{4bf_\pi^2}m_\pi+\Od(m_\pi^2\log m_\pi),\nonumber\\
T_{c2}&=&T_c+\frac{a T_{0ch}^2}{4bf_\pi^2}m_\pi+\Od(m_\pi^2\log m_\pi)=T_0-\frac{d_0m_K^2+d_1T_{0ch}^2}{4bf_\pi^2m_K^2}T_{0ch}m_\pi^2+\Od(m_\pi^3),\nonumber\\
T_{c3}&=&T_{0}+\frac{c_1T_{0ch}^2-c_0m_K^2}{4bf_\pi^2m_K^2}T_{0ch}m_\pi^2+\Od(m_\pi^3),
\label{tempchiral}
\end{eqnarray}
which is consistent with the numerical results showed in Fig.~\ref{fig:chiraltc} and with the $T_{c3}>T_0>T_{c2}>T_c$ hierarchy. 
In addition, the gap between the $U(1)_A$ pseudocritical temperatures $T_{c3}$ and $T_{c2}$ is $\Od(m_\pi^2)$, which is also the gap between them and $T_0$. 
On the contrary, the gap between $T_0$, $T_{c2}$ or $T_{c2}$ and the $O(4)$ $T_c$ is $\Od(m_\pi)$, i.e., larger in the chiral limit expansion.  

The chiral expansion of the $U(3)$ ChPT results is also particularly useful to disentangle the behavior of the connected and disconnected parts of the scalar susceptibility, which we have discussed in a general context in Section~\ref{sec:ib}. The ChPT expansion, by construction, is not able to generate a peak for the scalar susceptibility as $T\rightarrow T_c$. However, we can learn about the critical behavior of the different susceptibilities involved by examining their infrared (IR) chiral limit $m_\pi\rightarrow 0^+$ behavior, for which ChPT does capture the expected behavior for condensates and susceptibilities~\cite{Smilga:1995qf,Ejiri:2009ac}. 

Thus, consider the behavior of the different susceptibilities involved in the relation~\eqref{chi5vschidisc} in the chiral limit at $O(4)$ and $O(4)\times U(1)_A$ restoration, i.e., at  $T_c$ and  $T_{c3}$. 
On the one hand, we have at $T=T_c$
\begin{eqnarray}
\tilde\chi_S^{dis}(T_c)&=&a\left(B_0^r\right)^2\frac{T_c}{4m_\pi}+\Od(\log m_\pi),\nonumber\\
\chi_{5,disc}(T_c)&=&\tilde\chi_S^{dis}(T_c)+\frac{1}{4}\left(B_0^r\right)^2\left[\frac{(c_1+d_1)T_c^2}{m_K^2}+d_0-c_0\right]+\Od(m_\pi),\nonumber\\
\chi_S^\delta (T_c)&=&\left(B_0^r\right)^2\left(d_0+d_1\frac{T_c^2}{m_K^2}\right)+\Od(m_\pi),\nonumber\\
\chi_P^{ll}(T_c)=&=&\left(B_0^r\right)^2\left(c_0-c_1\frac{T_c^2}{m_K^2}\right)+\Od(m_\pi),
\label{IRTc}
\end{eqnarray}
which stem from~\eqref{suschiral} and~\eqref{tempchiral} with $\tilde\chi_S^{dis}=\frac{1}{4}\left[\tilde\chi_S^{ll}-\chi_S^\delta\right]$ according to the discussion in Section~\ref{sec:ib}. 
Therefore, at $T_c$ the IR divergent behavior of $\chi_{5,disc}$ in the l.h.s. of~\eqref{chi5vschidisc} is carried entirely by $\tilde\chi_S^{dis}$ in the r.h.s.. Note that the second term in the r.h.s. of~\eqref{chi5vschidisc} vanishes by definition at $T_c$ and the third term in the r.h.s. is regular in the IR limit. 

On the other hand, at $T=T_{c3}$ one finds
\begin{eqnarray}
\tilde\chi_S^{dis}(T_{c3})&=& a\left(B_0^r\right)^2\frac{T_{c3}}{4m_\pi}+\Od(\log m_\pi),\nonumber\\
\tilde\chi_S^{ll}(T_{c3})&=&4\tilde\chi_S^{dis}(T_{c3})+\chi_S^\delta (T_{c3}) +\Od(m_\pi),\nonumber\\
\chi_S^\delta (T_{c3})&=&\left(B_0^r\right)^2\left(d_0+d_1\frac{T_{c3}^2}{m_K^2}\right)+\Od(m_\pi).
\label{IRTc3}
\end{eqnarray}
Note that $T_{c3}$ is defined as the temperature for which $\chi_{5,disc}(T_{c3})=\chi_P^\pi (T_{c3})-\chi_P^{ll}(T_{c3})=0$.
This vanishing is compatible with the fact that  $\tilde\chi_S^{dis}(T_{c3})$ in the r.h.s of~\eqref{chi5vschidisc} is IR divergent, as given by~\eqref{IRTc3}. 
Namely, such divergence is exactly cancelled by that of $-\tilde\chi_S^{ll}(T_{c3})/4$. The remaining terms in~\eqref{chi5vschidisc} are IR regular and their sum vanishes exactly.

\begin{figure}
\centerline{\includegraphics[width=14cm]{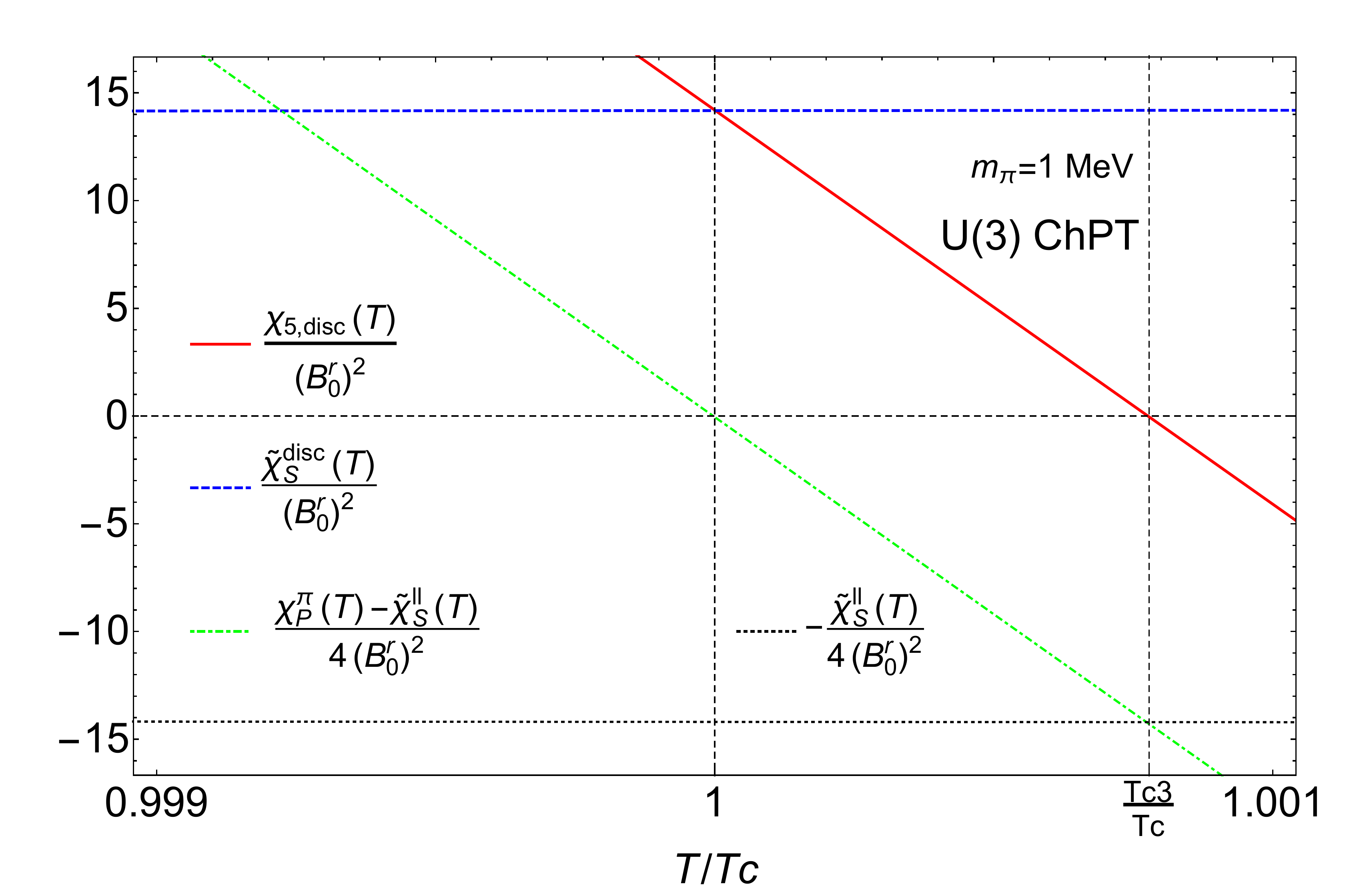}}
\caption{Behavior of susceptibilities in the decomposition~\eqref{chi5vschidisc} close to the chiral limit in $U(3)$ ChPT.}
\label{fig:susnearchiral}
\end{figure}

As a summary, it is perfectly compatible from a ChPT point of view  to have a divergent $\tilde\chi_S^{dis}$ and a vanishing $\chi_{5,disc}$ at $T_{c3}$ while both diverge at $T_c$, with $T_{c3}-T_c=\Od(m_\pi)$. 
These features can be appreciated in Fig.~\ref{fig:susnearchiral}, where we plot those susceptibilities very close to the chiral limit. At $T=T_c$, $\chi_P^\pi-\tilde\chi_S^{ll}$ vanishes while $\chi_{5,disc}$ and $\tilde\chi_S^{dis}$  are both large and of the same order, which arises from their $1/m_\pi$ behavior (compare with the typical numerical values of susceptibilities  in the physical case in Fig.~\ref{fig:foursuschpt}). At $T=T_{c3}$, $\chi_{5,disc}$ vanishes and the large positive value of $\tilde\chi_{dis}$ is compensated by the large negative contribution of -$\tilde\chi_S^{ll}/4$, as discussed above. 

\begin{figure}
\centerline{\includegraphics[width=14cm]{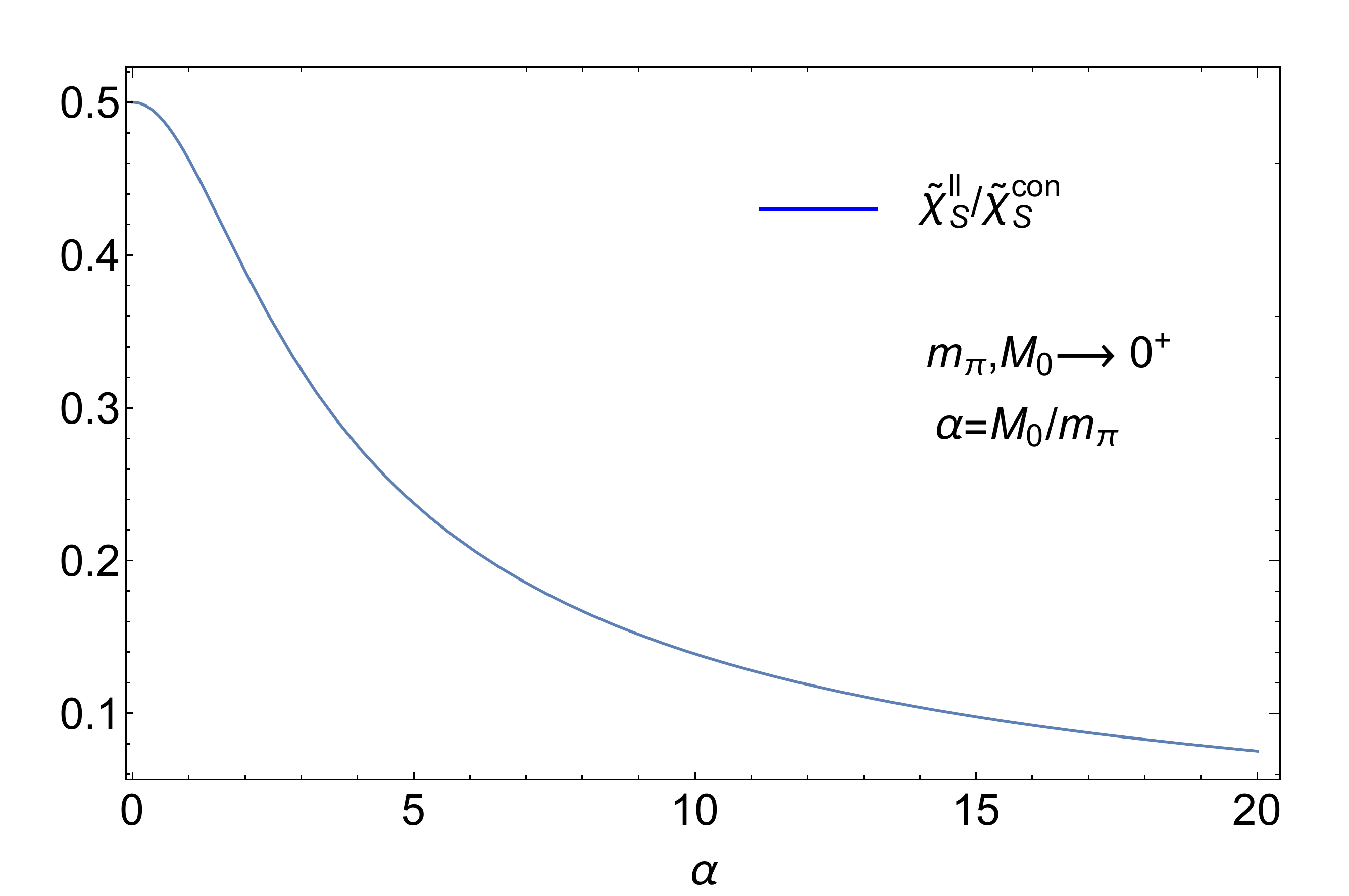}}
\caption{Ratio of connected to total scalar susceptibility in the combined limits $M_0\rightarrow 0^+$, $m_\pi\rightarrow 0^+$ with $\alpha=M_0/m_\pi$.}
\label{fig:alphafun}
\end{figure}

In the above discussion, the connected susceptibility, i.e., $\tilde\chi_S^{con}=\chi_S^\delta/2$, remains regular in the chiral limit. 
Nevertheless, as already discussed in Section~\ref{sec:ib}, general arguments indicate that $\tilde\chi_S^{con}$ could actually peak near $U(1)_A$ restoration. 
A hint of that behavior can be seen also in $U(3)$ ChPT by taking simultaneously the limits $m_\pi\rightarrow 0^+$ and $M_0\rightarrow 0^+$.
Note that $M_0$ is the anomalous part of the $\eta'$ mass, which should vanish in a $U(1)_A$ restoring scenario. 
The contributions to $\chi_\delta$ include mixed loop terms of the form~\eqref{mixedterms} with $i=\pi,j=\eta$.
In the $M_0\rightarrow 0^+$ limit, we have $m_\eta\rightarrow m_\pi^+$, leading to
\begin{equation}
\lim_{m_\eta\rightarrow m_\pi}\frac{g_1(m_\eta,T)-g_1(m_\pi,T)}{m_\eta^2-m_\pi^2}=\frac{1}{2m_\pi}\frac{d}{d m_\pi}g_1(m_\pi,T)=\frac{d}{d m_\pi^2} g_1(m_\pi,T)=-g_2(m_\pi,T).
\end{equation}
which, according to~\eqref{g1g2exp}, generates an additional IR divergent term not present in the $m_\pi\rightarrow 0^+$ for a fixed $m_\eta$. 
In more detail, in the $m_\pi\rightarrow 0^+$ and $M_0\rightarrow 0^+$ limit we obtain
\begin{eqnarray}
\tilde\chi_S^{ll}(T)&\stackrel{M_0, m_\pi \rightarrow 0^+}{\longrightarrow}& \left(B_0^r\right)^2\frac{3+\frac{1}{\sqrt{1+\frac{2 \alpha^2}{3}}}}{4 \pi}  \frac{T}{m_\pi}
+\Od(\log m_\pi),\nonumber\\
\tilde\chi_S^{con}&\stackrel{M_0, m_\pi \rightarrow 0^+}{\longrightarrow}&\left(B_0^r\right)^2         \frac{\sqrt{6 \alpha^2+9}-3}{2 \pi  \alpha^2} \frac{T}{m_\pi}+\Od(\log m_\pi),
\label{IRandM0}
\end{eqnarray}
with $\alpha=M_0/m_\pi$. We see that the connected scalar susceptibility above contains an IR divergent part in this combined limit, whose strength is parameterized by $\alpha$. 
On the one hand, taking $\alpha\rightarrow\infty$ we recover in~\eqref{IRandM0} the results given in~\eqref{suschiral}, corresponding to $m_\pi\rightarrow 0^+$ and $M_0\neq 0$. 
On the other hand, the $\alpha\rightarrow 0^+$ limit would correspond to the maximum $U(A)_1$ restoration in this parameterization. 
In Fig.~\ref{fig:alphafun} we plot the ratio $\tilde\chi_S^{con}/\tilde\chi_S^{ll}$ at leading order in $T/m_\pi$ as a function of $\alpha$. 
We see that for $\alpha\rightarrow 0^+$ a maximum finite value of $1/2$ is reached for that ratio. 
For reference, the  value of $\alpha$ corresponding to the physical values of $m_\pi$ and $M_0$ is $\alpha\simeq 5.99$, which corresponds in Fig.~\ref{fig:alphafun} to $\tilde\chi_S^{con}/\tilde\chi_S^{ll}\simeq 0.21$. 

\begin{figure}
\centerline{\includegraphics[width=9cm]{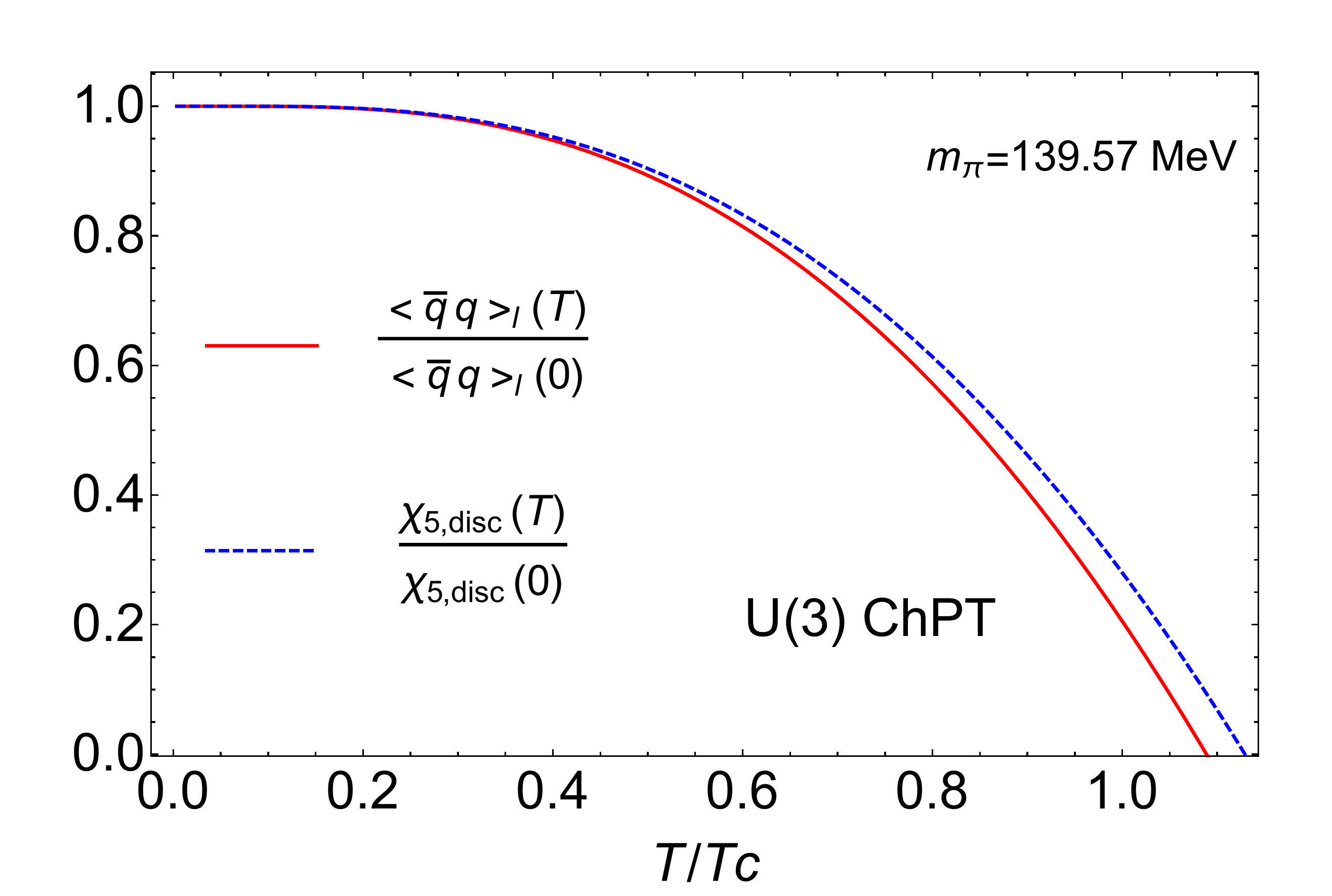}\includegraphics[width=9cm]{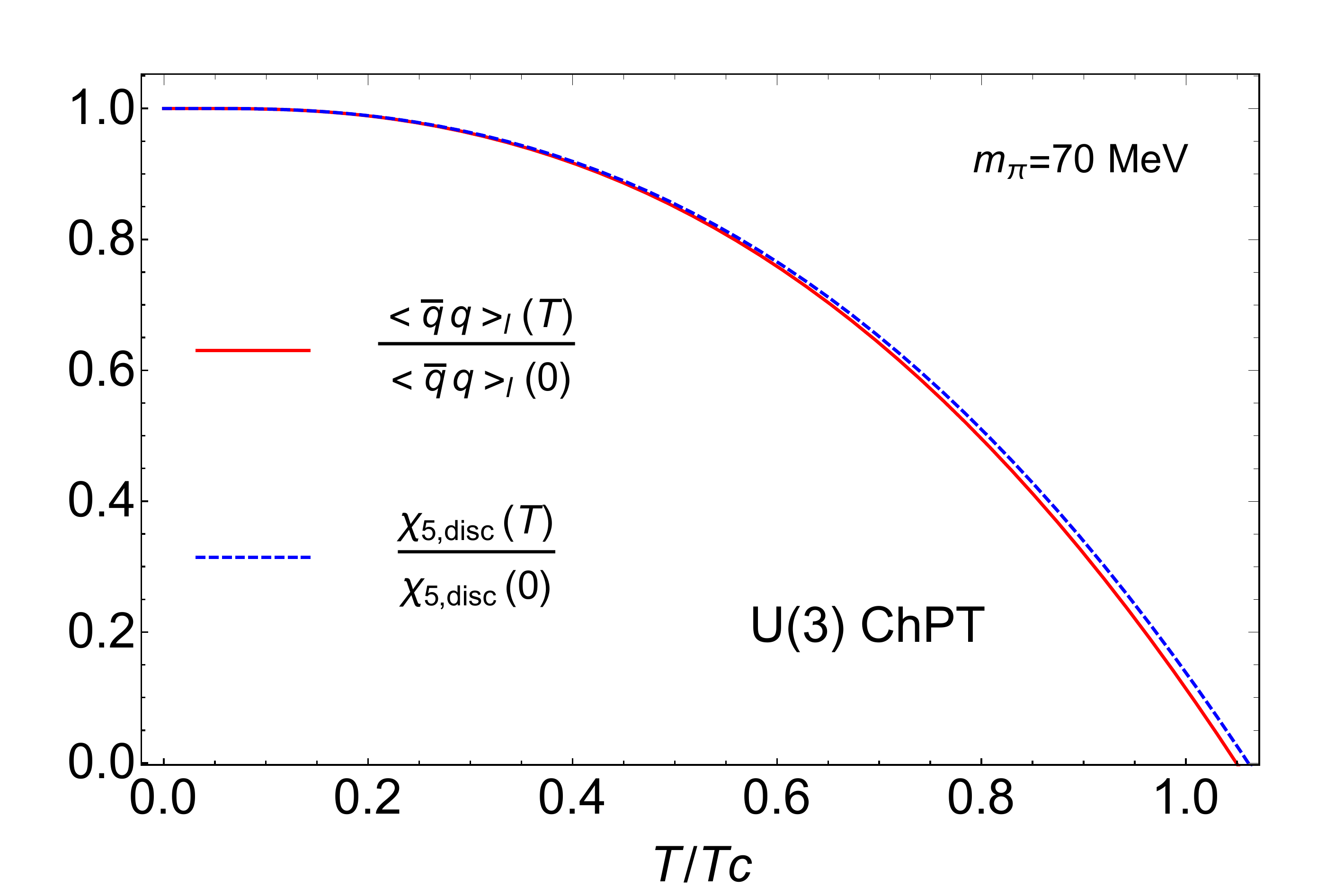}}
\centerline{\includegraphics[width=9cm]{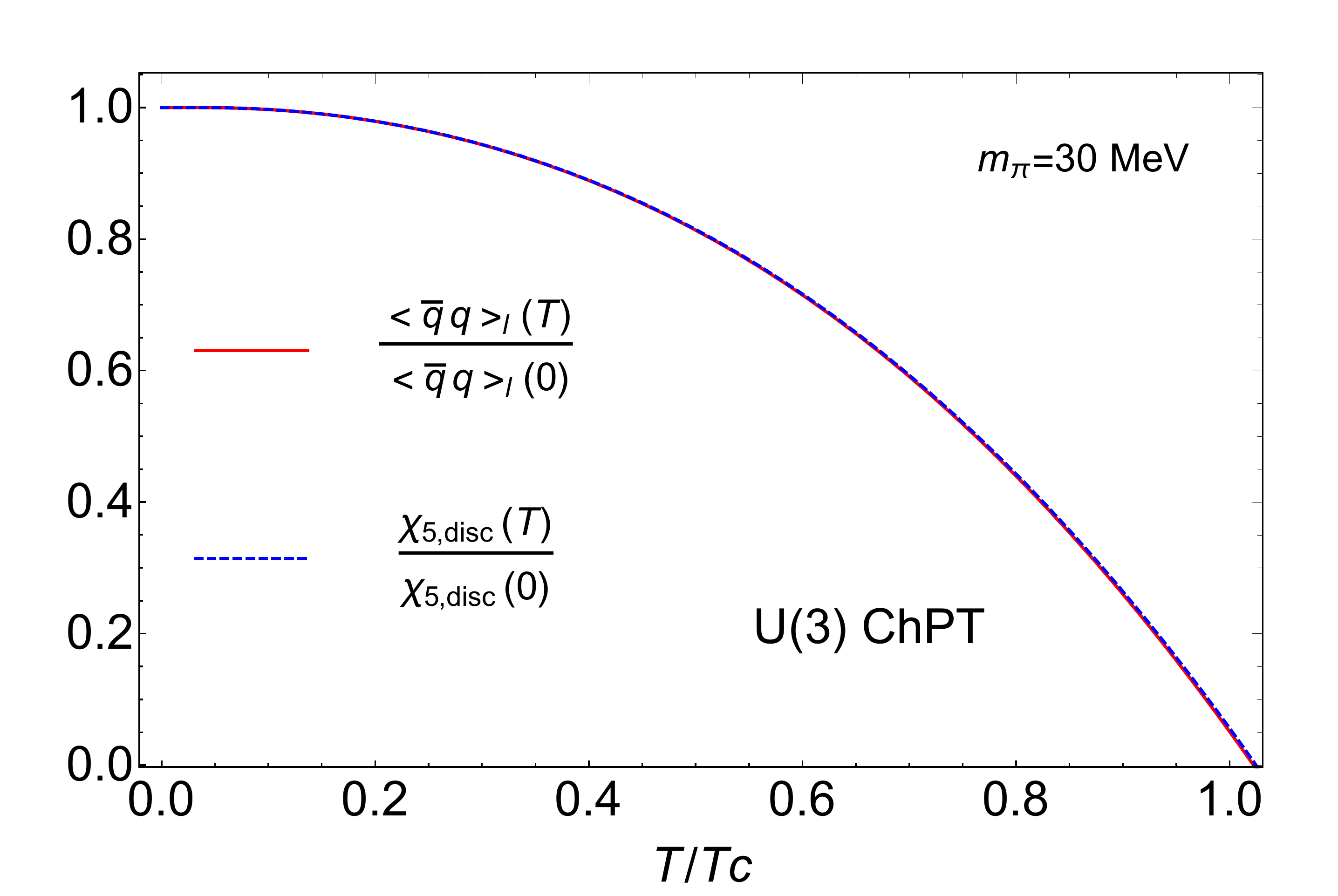}\includegraphics[width=9cm]{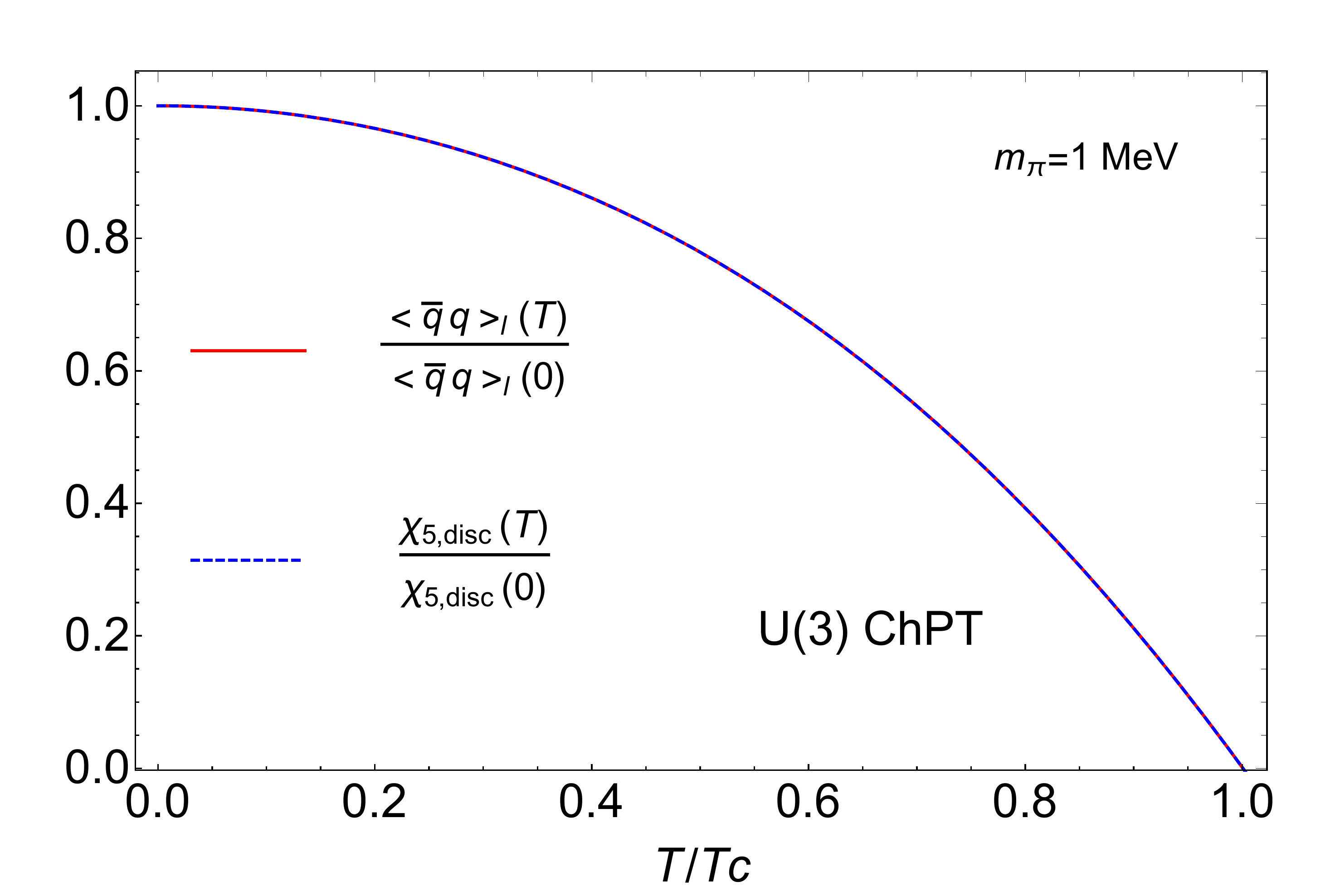}}
\caption{Comparison of temperature scaling of $\chi_{5,disc}$ and $\condl(T)$ for different values of the pion mass}
\label{fig:chi5vscond}
\end{figure}

Following the discussion in Section~\ref{sec:I01}, let us now compare the temperature scaling of $\chi_{5,disc}(T)$ and the light quark condensate $\condl (T)$. 
In Fig.~\ref{fig:chi5vscond} we plot $\chi_{5,disc}(T)/\chi_{5,disc}(0)$  and $\condl (T)/\condl (0)$ as the pion mass is reduced. 
It is clear that their temperature scaling is almost identical as the chiral limit is approached, consistently with~\cite{Azcoiti:2016zbi} and with our analysis in Section~\ref{sec:I01}. 
The reason can be understood again from the chiral limit expressions~\eqref{suschiral}. 
In the chiral limit the $\eta_l$ contribution $\chi_P^{ll}$ is parametrically negligible with respect to $\chi_P^\pi$, so that their difference given by $\chi_{5,disc}$ is dominated by $\chi_P^\pi$, which vanishes exactly like $\condl$ due to the WI~\eqref{wichip}.

\begin{figure}
\centerline{\includegraphics[width=14cm]{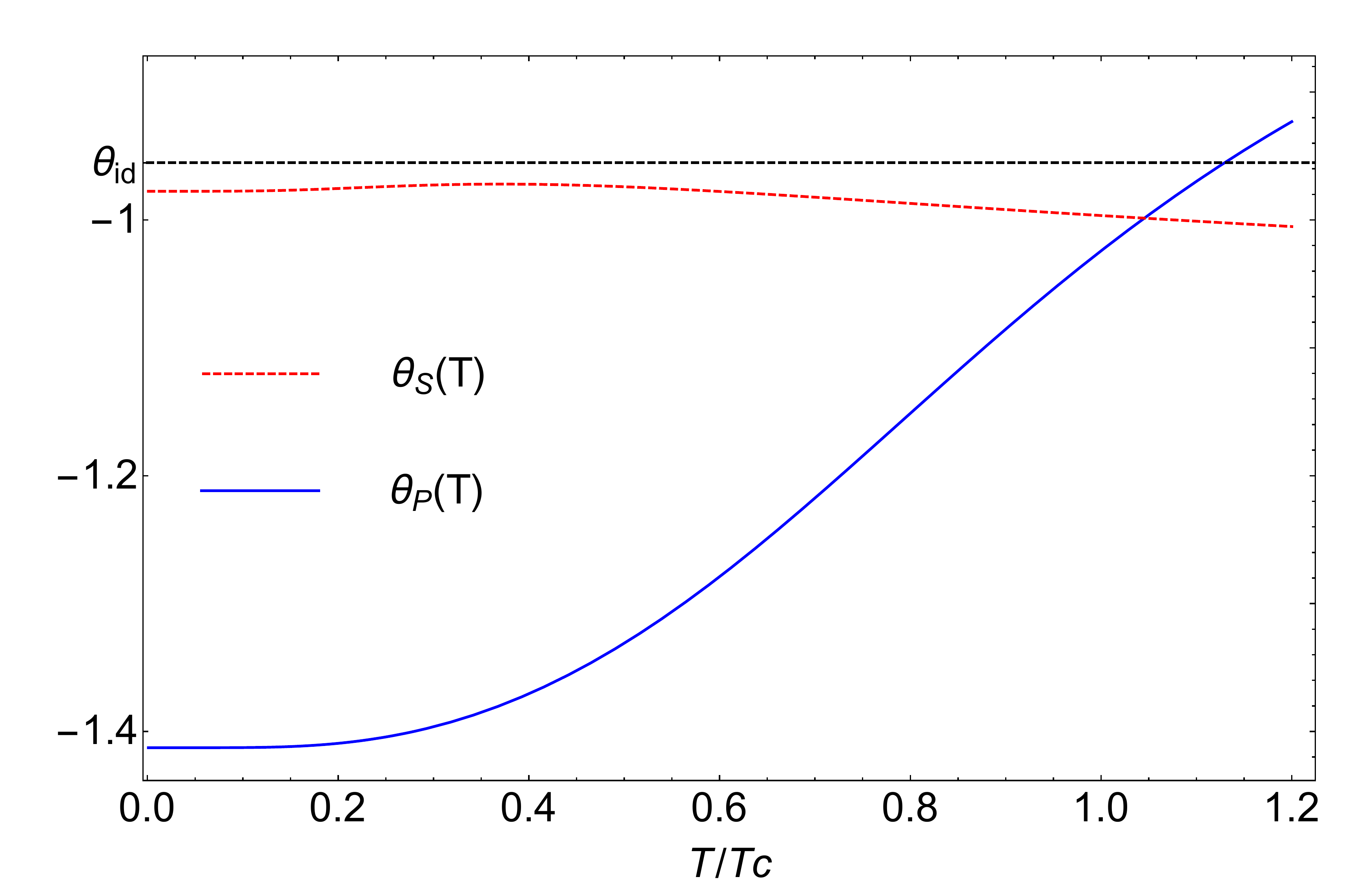}}
\caption{Temperature dependence of scalar and pseudoscalar mixing angles according to the definitions given in the main text.}
\label{fig:mixingangles}
\end{figure}

Finally, we will analyze the behavior of the scalar and pseudoscalar mixing angles. 
With the mixing angle defined through~\eqref{mixcond}, we solve for every $T$ the equations
\begin{equation}
\frac{1}{2}\left[\chi_{P,S}^{88}(T)-\chi_{P,S}^{00}(T)\right]\sin\left[2\theta_{P,S} (T)\right]  + \chi_{P,S}^{08}\cos \left[2\theta_{P,S}(T)\right]=0,
\end{equation} 
using the $U(3)$ ChPT expressions for the susceptibilities. The result is showed in Fig.~\ref{fig:mixingangles}. 
First, as commented in Section~\ref{sec:mixing}, the degeneration of the scalar and pseudoscalar mixing angles takes place at about $T\simeq 1.05 T_c$, i.e., around $O(4)\times U(1)_A$ degeneration. 
In addition, they coincide in a value close to the ideal mixing $\theta^{id}$, also consistently with the discussion in Section~\ref{sec:mixing}. 
In the case of $\theta_S$, the variation with respect to its $T=0$ value is small and close to ideal mixing. 
These findings are in fair agreement with the results in~\cite{Ishii:2015ira} obtained within the framework of the Polyakov-loop extended NJL model. 
Note that we do not see in this $U(3)$ ChPT analysis a region of vanishing mixing, since that would require a larger gap between $O(4)$ and $O(4)\times U(1)_A$ restoration. 

\section{Conclusions}
\label{sec:conc}
  
 In this work we have performed a detailed analysis of the correlators and susceptibilities corresponding to the scalar and pseudoscalar meson nonets, both from general arguments in terms of Ward Identities and from the model-independent description provided by $U(3)$ Chiral Perturbation Theory. Our main physical motivation has been the study of partners and patterns of chiral and $U(1)_A$ restoration. 
 
In particular, we have showed that in the limit of exact $O(4)$ restoration, understood in terms of $\delta-\eta$ partner degeneration, the WI analyzed yield also $O(4)\times U(1)_A$ restoration in terms of $\pi-\eta$ degeneration, i.e., from the vanishing of $\chi_{5,disc}$. Our analysis also provides a connection between $\chi_{5,disc}$ and the topological susceptibility $\chi_{top}$, which is defined from the correlator of the anomaly operator.
The results we obtain using ChPT are consistent with this analysis. Namely, one finds that the pseudo-critical temperatures for restoration of $O(4)$ and $O(4)\times U_A(1)$ tend to coincide in the chiral limit. 
In the real physical world with massive quarks, our conclusions agree with $N_f=2$ lattice results for partner degeneration. 
The large gap between $O(4)$ and $O(4)\times U(1)_A$ partner degeneration observed in $N_f=2+1$ simulations can be explained by the distortion in $\delta-\eta$ degeneration, presumably induced by strange quark mass effects. 
The large NLO corrections for the $\eta_l$ susceptibility that we obtain within $U(3)$ ChPT support this conclusion. 
  
In addition, including isospin breaking $m_u\neq m_d$ effects, we have recovered the formal connection of the $\delta$ and $\sigma$ susceptibilities with the connected and disconnected scalar ones, customarily measured in lattice analysis. The behavior of the connected and disconnected contributions to the scalar susceptibility have been studied within ChPT near $O(4)$ and $O(4)$ and $O(4)\times U(1)_A$ restoration. 
In that context, we have shown that a vanishing $\chi_{5,disc}$ at $O(4)\times U(1)_A$ restoration is compatible with a divergent $\chi_S^{disc}$. 
Moreover, the ChPT behavior for a vanishing $M_0$ (the anomalous part of the $\eta'$ mass) is a hint towards a possible peaking behavior of the connected $\chi_S^{con}$.

Regarding scalar and pseudoscalar mixing angles, our analysis shows that the WI are consistent with $\theta_P\sim \theta_S\sim\theta^{id}$ degeneration around $O(4)\times U(1)_A$ restoration, where $\theta^{id}$ is the ideal mixing angle.That conclusion is supported also by the $U(3)$ ChPT analysis, where $\theta_S$ remains close to ideal mixing for all temperatures, consistently with recent analyses. 
In the $N_f=2+1$ lattice data, an intermediate range between $O(4)$ and $O(4)\times U(1)_A$ restoration, compatible with vanishing pseudoscalar mixing is present.  

Our analysis shows also that in the $I=1/2$ sector, the $K$ and $\kappa$ states degenerate both at exact $O(4)$ and $U(1)_A$ restoration. 
Moreover, the degree of degeneracy of these two patterns is directly determined by the subtracted condensate $\Delta_{l,s}$ measured in the lattice. 
These results are confirmed also within the $U(3)$ ChPT analysis. In addition,  we have also showed in this sector that the temperature behavior of the screening mass in the $\kappa$ channel measured in the lattice can be explained with the corresponding WI relating $\chi_S^\kappa$ with the difference of light and strange quark condensates, which we have checked in ChPT. 
Such analysis extends a previous work for the $\pi,K,\eta$ channels.  We have also showed that the four channels can be simultaneously described with a two-parameter fit. 
  
  Our $U(3)$ ChPT analysis allows one to obtain  all the nonet scalar susceptibilities up to NNLO in the chiral power counting for finite temperature, thus completing  previous calculations of the pseudoscalar ones. The explicit expressions for those scalar susceptibilities are also provided here. 
  
In addition, we have discussed additional WI relating two and three-point functions, which may become useful to relate $O(4)$ and $U(1)_A$ partner degeneration quantities with meson vertices and scattering amplitudes. 
A detailed analysis of those WI is left for future investigation. 
 
As a summary, our study provides new theoretical insight for the understanding of the nature of the chiral and $U(1)_A$ transitions in terms of the degeneration of the meson nonet states, which is meant to be useful for lattice, phenomenological and experimental analyses. 
The picture emerging both from a general Ward Identity framework and from ChPT is robust and provides model-independent conclusions that could guide future work on this subject.
      
\section*{Acknowledgments}
We thank Z. H. Guo, S. Sharma, O. Phillipsen and F.Karsch for helpful discussions.  A.G.N is very grateful to the AEC and the ITP in Bern, and to the Theoretical High Energy Physics group of the Bielefeld  University, for their hospitality and financial support.  Work partially supported by  research contracts FPA2014-53375-C2-2-P, FPA2016-75654-C2-2-P  (spanish ``Ministerio de Econom\'{\i}a y Competitividad") and the Swiss National Science Foundation.

\appendix

\section{ChPT results} 
\label{sec:app} 

In this appendix we provide the explicit $U(3)$ ChPT expressions for the scalar susceptibilities $\tilde\chi_S^{ll}(T)$, $\tilde\chi_S^{ss}(T)$, $\chi_S^{ls}(T)$, $\chi_S^{\delta}(T)$ and $\chi_S^\kappa(T)$.
Up to NNLO in the $\delta$ expansion, one finds
\begin{align}
\chi_S^{ll}=&4B_0^{r\;2}\left(-3\nu_\pi-\nu_K -{1\over9}\left(4-c_\theta^2(3-7s_\theta^2)-4\sqrt 2c_\theta s_\theta(1+s_\theta^2)\right)\nu_\eta-{1\over9}\left(4+s_\theta^2(4-7s_\theta^2)-4\sqrt 2c_\theta s_\theta(-2+s_\theta^2)\right)\nu_{\eta'}\right.\nonumber\\
&+\frac{2F^2}{9}\frac{\left(2c_\theta^4-2\sqrt2c_\theta^3s_\theta-3c_\theta^2s_\theta^2+2\sqrt2c_\theta s_\theta^3+2s_\theta^4\right)\left(\mu_\eta-\mu_{\eta'}\right)}{m_{0\eta'}^2-m_{0\eta}^2}+4\left(8L_6^r+2L_8^r+H_2^r+{8\over 3}C_{19}^r(4m_{0K}^2-m_{0\pi}^2)\right)\Bigg),\nonumber\\\nonumber\\
\chi_S^{ss}=&4B_0^{r\;2}\left(-\nu_K -{1\over9}\left(1+c_\theta^2(3+7s_\theta^2)+4\sqrt 2c_\theta s_\theta(2-s_\theta^2)\right)\nu_\eta-{1\over9}\left(1+s_\theta^2(10-7s_\theta^2)-4\sqrt 2c_\theta s_\theta(1+s_\theta^2)\right)\nu_{\eta'}\right.\nonumber\\
&+\frac{2F^2}{9}\frac{\left(2c_\theta^4-2\sqrt2c_\theta^3s_\theta-3c_\theta^2s_\theta^2+2\sqrt2c_\theta s_\theta^3+2s_\theta^4\right)\left(\mu_\eta-\mu_{\eta'}\right)}{m_{0\eta'}^2-m_{0\eta}^2}+2\left(4L_6^r+2L_8^r+H_2^r+{16\over 3}C_{19}^r(4m_{0K}^2-m_{0\pi}^2)\right)\Bigg),\nonumber\\\nonumber\\
\chi_S^{ls}=&4B_0^{r\;2}\left(-\nu_K -{1\over9}\left(2-7c_\theta^2s_\theta^2-2\sqrt 2c_\theta s_\theta(1-2s_\theta^2)\right)\nu_\eta-{1\over9}\left(1-7s_\theta^2(1-s_\theta^2)-2\sqrt 2c_\theta s_\theta(1-2s_\theta^2)\right)\nu_{\eta'}\right.\nonumber\\
&+\frac{2F^2}{9}\frac{\left(2c_\theta^4-2\sqrt2c_\theta^3s_\theta-3c_\theta^2s_\theta^2+2\sqrt2c_\theta s_\theta^3+2s_\theta^4\right)\left(\mu_\eta-\mu_{\eta'}\right)}{m_{0\eta'}^2-m_{0\eta}^2}+16\left(L_6^r-{2\over 3}C_{19}^r(4m_{0K}^2-m_{0\pi}^2)\right)\Bigg),\nonumber\\\nonumber\\
\chi_S^\delta=&4B_0^{r\;2}\left(-\nu_K+\frac{2F^2}{3}\frac{\left(c_\theta^2-2\sqrt 2c_\theta s_\theta +2s_\theta^2\right)\left(\mu_\pi-\mu_\eta\right)}{m_{0\eta}^2-m_{0\pi}^2}+\frac{2F^2}{3}\frac{\left(2c_\theta^2+2\sqrt 2c_\theta s_\theta +s_\theta^2\right)\left(\mu_\pi-\mu_{\eta'}\right)}{m_{0\eta'}^2-m_{0\pi}^2}\right.\nonumber\\
&+4(2L^r_8+H^r_2+24C_{19}^rm_{0\pi}^2)\Bigg),\nonumber\\\nonumber\\
\chi_S^\kappa=&2B_0^{r\;2}\left(\frac{3F^2(\mu_\pi-\mu_K)}{m_{0K}^2-m_{0\pi}^2}+\frac{F^2}{3}\frac{\left(c_\theta^2+4\sqrt 2c_\theta s_\theta +8s_\theta^2\right)\left(\mu_K-\mu_\eta\right)}{m_{0\eta}^2-m_{0K}^2}+\frac{F^2}{3}\frac{\left(8c_\theta^2-4\sqrt 2c_\theta s_\theta+s_\theta^2\right)\left(\mu_K-\mu_{\eta'}\right)}{m_{0\eta'}^2-m_{0K}^2}\right.\nonumber\\
&+8\left(2L_8^r+H_2^r+24C_{19}^rm_{0K}^2\right)\Bigg),
\end{align}
where the $T$-dependent loop functions $\mu_i$ and $\nu_i$ are defined in~\eqref{mudef} and~\eqref{nudef}, respectively.

In addition, $m_{0\pi}=2B^r_0 \hat m$ and $m_{0K}=B^r_0(\hat m+m_s)$ stand for the LO pion and kaon masses, 
whereas the LO $\eta$ and $\eta'$ masses are given by:
\begin{align}
m_{0\eta}^2=&\frac{M_0^2}{2}+m_{0K}^2-\frac{\sqrt{M_0^4-{4M_0^2\Delta^2\over 3}+4\Delta^4}}{2},\\
m_{0\eta'}^2=&\frac{M_0^2}{2}+m_{0K}^2+\frac{\sqrt{M_0^4-{4M_0^2\Delta^2\over 3}+4\Delta^4}}{2},
\end{align}
with $\Delta^2=m_{0K}^2-m_{0\pi}^2$ and $M_0$ the anomalous part of the $\eta_0$ mass.

Finally, $c_\theta\equiv\cos\theta_P$ and $s_\theta\equiv\sin\theta_P$, $\theta_P$ is the $\eta-\eta'$ mixing angle defined in~\eqref{mix}, which to LO reads
\begin{equation}
\sin\theta_P=-\left(1+\frac{\left(3M_0^2-2\Delta^2+\sqrt{9M_0^4-12M_0^2\Delta^2+36\Delta^4}\right)^2}{32 \Delta^4}\right)^{-1/2}.
\end{equation}

\end{document}